\documentclass[twocolumn]{aastex62}

\newcommand{\RSun}{\ensuremath{R_\mathrm{S}}}

\graphicspath{{./}{figures/}}

\shorttitle{Sample article}
\shortauthors{Sarkar et al.}

\begin{document}

\title{\textbf {\large Studying the spheromak rotation in data-constrained CME modelling with EUHFORIA and assessing its effect on the $B_z$ prediction}}

\correspondingauthor{Ranadeep Sarkar}
\email{ranadeep.sarkar@helsinki.fi}

\author{Ranadeep Sarkar}

\author{Jens Pomoell}

\author{Emilia Kilpua}

\author{Eleanna Asvestari}
\affiliation{Department of Physics, University of Helsinki, Helsinki, Finland} 

\author{Nicolas Wijsen}
\affiliation{NASA Goddard Space Flight Center, Greenbelt, MD 20771, USA}
\affiliation{Department of Astronomy, University of Maryland College Park, MD 20742, USA}
\author{Anwesha Maharana}
\affiliation{Centre for Mathematical Plasma Astrophysics, KU Leuven, Leuven, Belgium}
\affiliation{Royal Observatory of Belgium, 1180 Uccle, Belgium}
\author{Stefaan Poedts}
\affiliation{Centre for mathematical Plasma Astrophysics (CmPA)/Dept.\ of Mathematics, KU Leuven, 3001 Leuven, Belgium}
\affiliation{Institute of Physics, University of Curie-Sk{\l}odowska,  
	ul.\ Radziszewskiego 10, 20-031 Lublin, Poland}

\begin{abstract}

A key challenge in space weather forecasting is accurately predicting the magnetic field topology of interplanetary coronal mass ejections (ICMEs), specifically the north-south magnetic field component ($B_z$) for Earth-directed CMEs. Heliospheric MHD models typically use spheromaks to represent the magnetic structure of CMEs. However, when inserted into the ambient interplanetary magnetic field, spheromaks can experience a phenomenon reminiscent of the condition known as the ``spheromak tilting instability", causing its magnetic axis to rotate. From the perspective of space weather forecasting, it is crucial to understand the effect of this rotation on predicting $B_z$ at 1~au while implementing the spheromak model for realistic event studies. In this work, we study this by modelling a CME event on 2013 April 11 using the “EUropean Heliospheric FORecasting Information Asset” (EUHFORIA). Our results show that a significant spheromak rotation up to 90$^\circ$ has occurred by the time it reaches 1~au, while the majority of this rotation occurs below 0.3~au. This total rotation resulted in poor predicted magnetic field topology of the ICME at 1~au. To address this issue, we further investigated the influence of spheromak density on mitigating rotation. The results show that the spheromak rotation is less for higher densities. Importantly, we observe a substantial reduction in the uncertainties associated with predicting $B_z$ when there is minimal spheromak rotation. Therefore, we conclude that spheromak rotation adversely affects $B_z$ prediction in the analyzed event, emphasizing the need for caution when employing spheromaks in global MHD models for space weather forecasting.

\end{abstract}

\keywords{Solar coronal mass ejections; Magnetohydrodynamical simulations; Space weather}

\section{Introduction} \label{sec:intro}
Coronal mass ejections (CMEs) are one of the major sources for space-weather disturbances. If the magnetic field inside an Earth-directed CME, or inside its associated sheath region, has a southward-directed north-south magnetic field component ($B_z$), then it interacts effectively with the Earth’s magnetosphere, leading to severe geomagnetic storms depending on the strength of $B_z$ \citep{Wilson_1987,tsurutani_1988,Gonzalez_1999,Huttunen_2005, Gopalswamy_2008}. Therefore, it is crucial to predict the strength and direction of $B_z$ inside Earth-impacting interplanetary CMEs (ICMEs) in order to forecast their geo-effectiveness. Since the magnetic field of CMEs cannot reliably be measured remotely, and direct in-situ measurements of Earth-impacting ICMEs are routinely available only very close to our planet, modelling of CME magnetic properties using near-Sun observational proxies is paramount.

The state-of-the-art global heliospheric MHD models typically implement the axi-symmetric spheromak or modified spheromak configurations to characterize the magnetic structure of a CME and simulate its evolution from Sun-to-Earth \citep{1998Vandas,2002Vandas,Manchester_2004,Shiota2016,2019scolini, Asvestari1,Singh_2022}. In a recent study by \citet{Asvestari2} it was found that a spheromak injected into the inner heliospheric domain gradually rotates, due to the presence of a torque, until it reduces its magnetic potential energy. The rotation stops when its magnetic moment aligns with the ambient magnetic field. This behaviour is a reminiscent of the condition known as the ``spheromak tilting instability" \citep{rosenbluth_bussac_1979, bellan_fundamentals_2000, bellan_2018, mehta_2020}. According to theory, the torque is larger in the presence of a strong ambient field. Taking into consideration that the magnetic field is indeed stronger near the Sun but drops with helio-distance as $1/r^{2}$, it is anticipated that the bulk of the spheromak rotation takes place near the Sun and the rotation rate drops with increasing helio-distance. Consistent to theory, in the modelling analysis of \citet{Asvestari2} it was found that the rotation rate was higher near the inner boundary at 0.1 astronomical unit (au) and dropped as the spheromak moved away from it. Despite the drop in the rotation rate, the overall rotation angle was still notable at large helio-distances. Observational evidence suggests that some CMEs show clear signatures of rotation in the corona \citep[e.g.,][]{Vourlidas_2011,Nieves-Chinchilla_2012,2015ApJKay}. CME rotation during its early phases of evolution is also noticed in simulations \citep{Torok_kliem_2003,2009ApJLynch}. However, such rotation mostly occurs in the low corona (below 0.05~au) and is less likely to happen at larger helio-distances in the heliosphere \citep{2015ApJKay}. Therefore, the presence of spheromak rotation even beyond 0.1~au 
%caused by manifestations of the tilting instability 
suggests that it may not be a realistic phenomena and thus can affect the performance of space-weather forecasting models that use spheromaks as a flux-rope.   

In the study of \citet{Asvestari2}, a strong or weak uni-directional (inward or outward) magnetic field in the ambient medium was employed to quantify the spheromak rotation. Studying this phenomenon in solar wind conditions representing a specific actual period of time will be an important step forward to assess its effect on space-weather forecasting. Event studies of past CMEs within the framework of global MHD models provide an excellent opportunity to compare the model results with in-situ observations at different helio-distances. Therefore, using event-based data-driven MHD simulations employing the spheromak model allows to build a quantitative understanding on how the spheromak rotation could affect the $B_z$ prediction at 1~au. 

There are several studies that have previously used spheromaks to study an observed CME event using data-driven MHD models \citep[e.g.,][]{Verbeke_2019, Asvestari1,2019scolini,Shiota2016}. However, a particular focus has not yet been made in any event-based study to address if the spheromak rotation has significant consequences for $B_z$ forecasts. In this work, we study a CME event on 2013 April 13 using the EUHFORIA model aiming to  understand the consequences of spheromak rotation in physics-based space-weather forecasting models. 

Previous studies with EUHFORIA use a default uniform mass density (1$\times10^{-18}$\;kg~m$^{-3}$) to specify the mass density inside the spheromak. However, a recent observational study shows that the average CME density at 0.1~au (inner boundary of global heliospheric MHD models) ranges within the order of 1$\times10^{-18}$\;kg m$^{-3}$ to 1$\times10^{-17}$\;kg m$^{-3}$ \citep{2021Mannuela}. Therefore, we also explore the effect of spheromak density in data-constrained CME modelling by performing a set of simulations with different spheromak densities which lie within the observational range. With this work, we address the following key scientific questions: 

\begin{enumerate}
    \item How does an observationally constrained spheromak evolves/rotates in a data-driven global MHD simulation? 
    \item  Until what distance from the Sun does the ambient magnetic field play a major role in the spheromak rotation?
    \item Does the spheromak density have a considerable effect on its heliospheric evolution?
    \item Does the spheromak evolution/rotation affect the prediction of $B_z$ at 1~au?
\end{enumerate}

To address these questions, we organise the paper as follows. First, we briefly discuss the CME event on 2013 April 11 in Section~\ref{section2}. The methods to obtain the background solar wind and the observational techniques to constrain the spheromak to mimic the associated CME structure for this event are described in Section~\ref{section3}. Based on our study of the spheromak evolution using EUHFORIA, we present the results in Section~\ref{section4}. Finally, we summarise these results in Section~\ref{section5} in the context of answering the science questions mentioned in Section~\ref{sec:intro}.

\begin{figure*}[t!]
\begin{center}
\includegraphics[width=\textwidth,clip=]{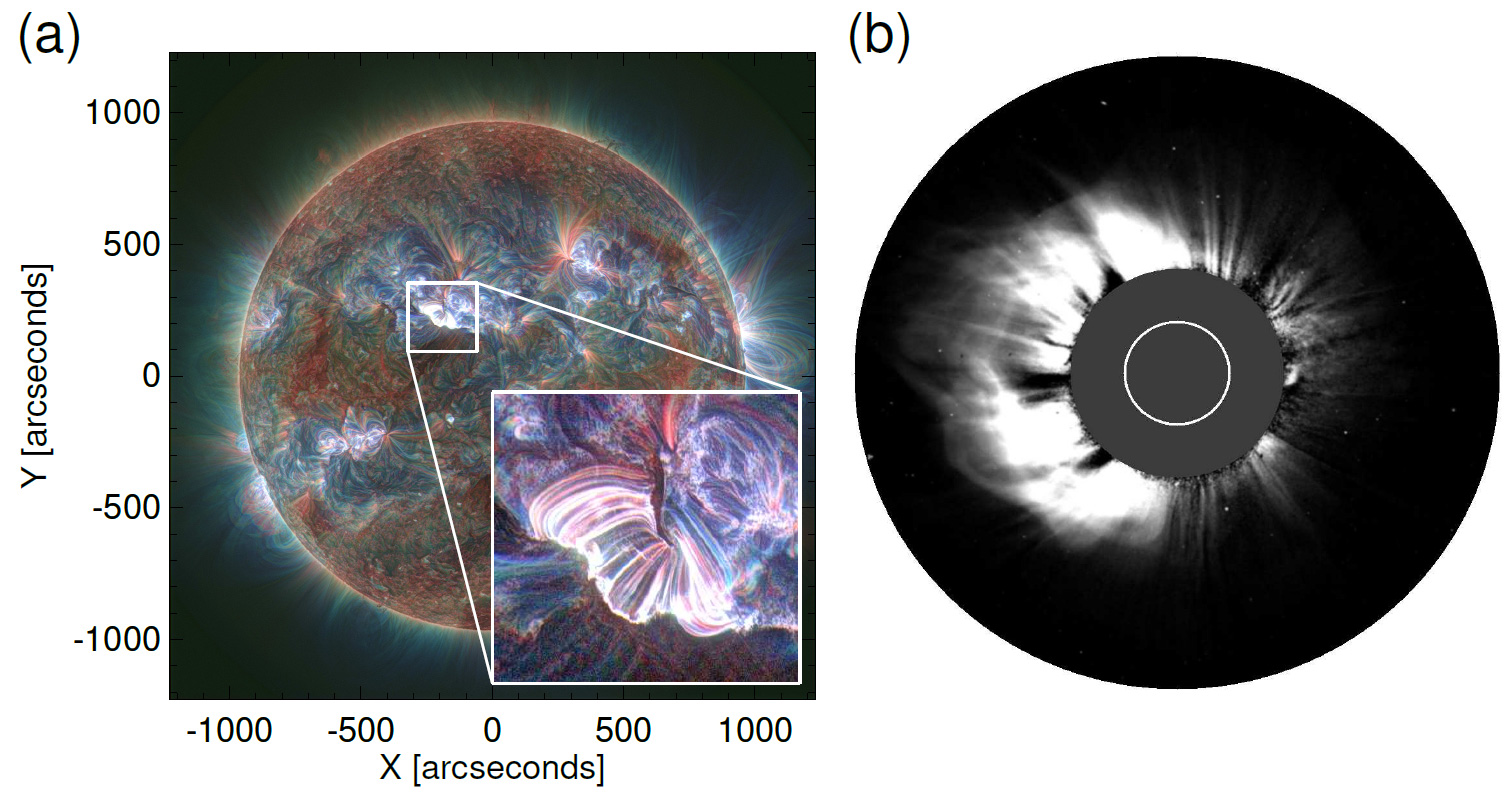}
\caption{Full-disk observation of the solar corona in EUV at 08:37~UT on 2013 April 11, as captured in the composite image constructed from the AIA 171 \AA\ (red), 211 \AA\ (green) and 193 \AA\ (blue) passbands (a). The inset in panel (a) depicts the morphology of the post eruption arcade associated with the M6.5 class flare. The white light structure of the associated CME observed in the LASCO C2 coronagraph at 07:54 UT on 2013 April 11 is shown in panel (b).}\label{aia_lasco}
\end{center}
\end{figure*}

\section{Overview of event}\label{section2}
We consider the CME event that erupted on 2013 April 11. This event has been studied extensively in previous works e.g., by \cite{Sarkar_2020} and \cite{Vemareddy_2015}. The CME appeared as a halo and was first observed at 07:24~UT in the field of view of LASCO/C2 coronagraph. The eruption was associated with an M6.5 class flare that occurred in NOAA active region (AR) 11719 at around 06:50 UT at the position of N07E13. Figure~\ref{aia_lasco} depicts the multi-wavelength observation of the CME source region in AIA passbands and the CME morphology in white-light as observed by LASCO. Based on the multi-spacecraft observations and reconstruction techniques, it was derived that the CME approximately propagates along the Sun-Earth line and was related to a well-defined magnetic cloud (MC) observed at 1~au \citep{Vemareddy_2015}. In situ observations from WIND reveal that the leading edge of the MC arrived at L1 on 2013 April 14 at around 17:00 UT \citep{Sarkar_2020}. Using the observations from heliospheric imagers (HIs) on board STEREO, the interplanetary propagation of the CME was studied in \citet{Vemareddy_2015} and it was found that the CME did not interact with any other CMEs during its propagation from Sun to Earth. The availability of cradle to grave observations of this event provides an excellent opportunity to model this event using near-Sun observational inputs and assess the model results at 1~au.

\section{Methods}\label{section3}

The underlying methodology of the modeling performed in this work is to first generate a steady-state ambient solar wind condition in the co-rotating frame, representing the large-scale heliospheric plasma environment prior to the observed eruption and then inject an observationally constrained spheromak into the medium to describe the propagating disturbance. The detailed methodology for the above mentioned steps are discussed below.

\subsection{Constructing the background solar wind}\label{bkg}
The background solar wind condition for the event under study is obtained using the modelling framework provided by EUHFORIA that consists of a semi-empirical coronal model - extending up to 0.1~au and a heliospheric MHD model - extending from 0.1~au to 2.0~au \citep{2018Pomoell}. Using the GONG synoptic magnetogram dated 2013 April 10 at 17:04 UT as initial input, we employ the Potential Field Source Surface (PFSS) and Schatten Current Sheet (SCS) \citep{mcgregor} extrapolation methods to get the coronal magnetic field up to 0.1~au. Afterwards, using the empirical formulations as described in \citet{2018Pomoell}, we generate all the associated MHD variables at the inner boundary (0.1~au) of the global MHD model. In order to account for the solar rotation, this solar wind map at 0.1~au is then rotated by an angle $\psi$ (default value is 10$^{\circ}$ as specified in EUHFORIA) to generate the final input for the heliospheric simulation. With this input boundary data, a steady-state solar wind solution is achieved by running the MHD simulation for a duration of 14 days. By changing the angle $\psi$ in the vicinity of its default value, we run a set of such simulations to optimize the background solar wind for this event and the optimized wind condition is achieved for $\psi = 17.7^{\circ}$. The optimization was done by comparing the simulation output with the observed in-situ profile of the high-speed stream ahead of the ICME at 1~au. More discussion on this is presented in section~\ref{insitu_assess}. This single relaxed optimized heliospheric state is then used as the starting point for all subsequent simulations in which a spheromak is injected following the description in the next section.\\

\begin{deluxetable}{ll}[!t]
\tablecaption{CME input parameters for simulation Run 1 \label{table1}}
\tablehead{
\colhead{Parameter} & \colhead{Value} 
}

\startdata
Insertion time & 2013-04-11T11:12:00 \\
Speed ($v_{radial}$) & 400\;km s$^{-1}$ \\
Longitude $\phi$ & 0$^{\circ}$ \\
Co-latitude $\theta$ & 0$^{\circ}$ \\
Radius & 7.2\;\RSun \\
Density & 1$\times10^{-18}$\;kg m$^{-3}$\\
Temperature & 0.8$\times10^{6}$\;K \\
Helicity & -1 (Left handed) \\
Tilt $\tau$ & -70$^{\circ}$ \\
Toroidal flux & $2.4\times10^{13}$ Wb \\
\enddata
\end{deluxetable}

\subsection{Constraining the spheromak from observations}
Information on several characteristics of the CME is needed to constrain the spheromak model that is injected at the inner boundary of the heliospheric model at 0.1~au. All the required parameters and the methods of how they are determined are described in the following subsections. We first describe the geometric parameters (tilt and radius) followed by the kinematic (speed and propagation direction) and magnetic (helicity sign and magnetic flux) parameters. Finally, we discuss the density and temperature to be used for the spheromak. The majority of the parameters that we use to constrain the model are based on the multi-wavelength and multi-spacecraft observational analysis reported in \cite{Sarkar_2020} and \cite{Vemareddy_2015}. Notably, these studies determine the CME parameters assuming that the underlying flux-rope structure is rooted in the Sun, which has a different geometry than that of the magnetically isolated spheromak. We emphasise below those differences and present the methods on how those CME parameters can be utilised to constrain a spheromak. The values used in the simulation runs are collected in Table~\ref{table1}.

\subsubsection{Tilt angle} The insertion tilt of the spheromak is defined as the orientation angle ($\tau$) of its symmetry axis measured clockwise from the meridional direction in the tangent plane to the inner spherical boundary ($r=0.1$~au) at the point ($\theta$, $\phi$) which corresponds to the co-latitude and longitude of the insertion direction of the spheromak \citep{Asvestari2,Verbeke2019}. For example, for an insertion direction of $\theta=0^{\circ}$, $\phi=0^{\circ}$ in the HEEQ coordinate system, $\tau=0^{ \circ}$ corresponds to a scenario where the symmetry axis of the spheromak is oriented along the Z-axis (along the solar rotation axis) so that its magnetic axis aligns along the Y-direction. Under the above-mentioned circumstance, the direction of the spheromak magnetic axis at the apex point (distant point from the Sun-center) on its magnetic axis curve points towards the positive Y-axis. 

\begin{figure}[t!]
\begin{center}
\includegraphics[width=.5\textwidth,clip=]{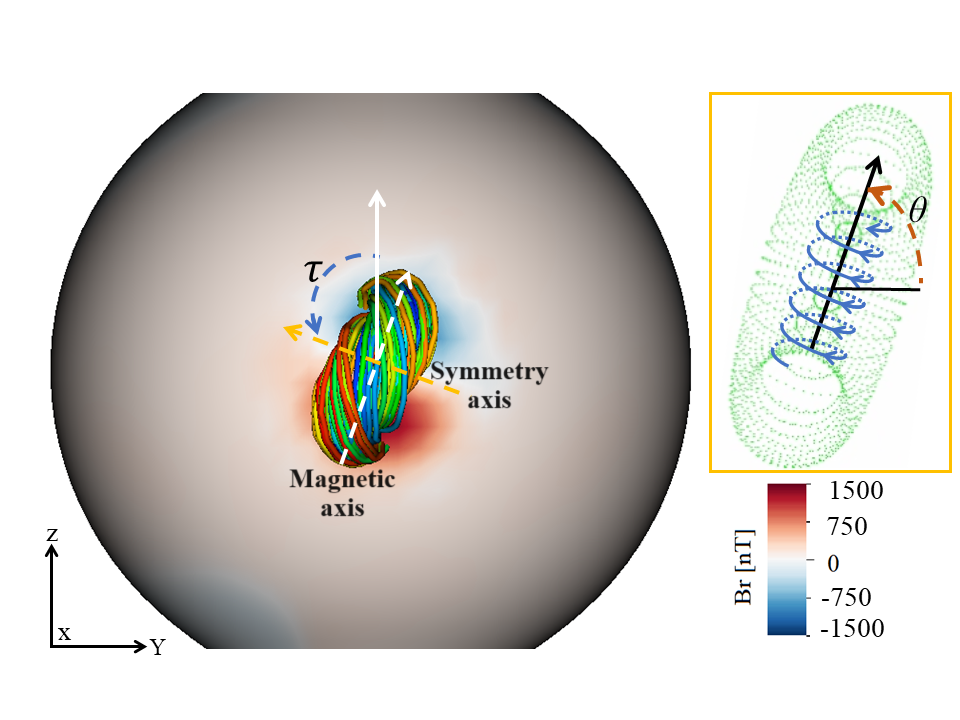}
\caption{The orientation of the spheromak during its insertion phase. The spherical surface denotes the lower boundary of the heliospheric domain in EUHFORIA at 0.1~au. The colour map represents the radial component of the magnetic field at this inner spherical boundary. The twisted field lines surrounding the magnetic axis of the spheromak are shown as they emerge from the inner boundary. The white and yellow dashed arrows approximately mark the directions of the magnetic and symmetry axis of the spheromak respectively. The white vertical arrow denotes the direction of the positive z-axis in the HEEQ system. The angle between the symmetry axis and the positive z-axis is marked as the spheromak tilt angle ($\tau$). The schematic picture shown in the inset depicts the orientation of the GCS mesh as well as the chirality (represented by blue helical curves rotating in an anti-clockwise direction) and axis direction (black arrow) of the flux-rope as estimated from remote-sensing observations. The GCS tilt angle as measured counterclockwise from the solar equator is marked as $\theta$.}\label{ini_tilt}
\end{center}
\end{figure}

On the other hand, the axis direction of a croissant-like flux-rope obtained from a morphological fit to the white-light observations is different from the tilt angle defined above. Observationally, the axial field direction of a croissant-like flux-rope is determined from the near-Sun magnetic proxies and geometrical tilt angle of the flux-rope obtained using the graduated cylindrical-shell (GCS) model \citep{gcs1,gcs2}. However, this axis orientation obtained from GCS refers to the orientation of the magnetic axis (relative to the solar equator) of a spheromak which is perpendicular to its symmetry axis \citep{Asvestari1}. Therefore, one will need to translate the flux-rope tilt extracted from observations with the GCS model to the insertion tilt of a spheromak in EUHFORIA.

Based on the near-Sun magnetic proxies (orientation of the magnetic connectivity between the two foot-points of the pre-eruptive sigmoid) and the GCS fitting, the axis orientation at the apex of the CME flux-rope under study is reported as northward with an angle of 70$^{\circ}$ at approximately 10 solar radii (\RSun) \citep{Sarkar_2020}. This angle ($\theta$) is measured counterclockwise from the solar equator as shown in the inset of Figure~\ref{ini_tilt}. As the most significant rotation of a CME mostly occurs close (below 10\;\RSun) to the Sun (see the Introduction), we use the axial direction of the CME obtained at $\approx$ 10\;\RSun\ as the input tilt angle ($\theta$) at 0.1~au, i.e., we assume that no further change of direction of the magnetic field structure occurs between 10 to 21.5\;\RSun.  

In order to estimate the associated angle $\tau$, we first transform the orientation angle ($\theta$=70$^{\circ}$) of the magnetic axis measured counterclockwise from the solar equator to that (20$^{\circ}$; 90$^{\circ}$-$\theta$) measured clockwise from the meridional direction. Notably, the symmetry axis (shown by the yellow dashed arrow in Figure~\ref{ini_tilt}) of a spheromak is 90$^{\circ}$ apart (counterclockwise) from the direction of its magnetic axis (shown by the white dashed arrow in Figure~\ref{ini_tilt}) as mentioned above. Therefore, applying a counterclockwise rotation of 90$^{\circ}$ to the orientation angle (20$^{\circ}$) of the magnetic axis of the spheromak, we determine the orientation angle of its symmetry axis as -70$^{\circ}$. We use this value (-70$^{\circ}$) as the input tilt angle ($\tau$) for the spheromak. Figure~\ref{ini_tilt} shows the orientation of the magnetic axis of the spheromak during its insertion phase which is consistent with the orientation (see the inset of Figure~\ref{ini_tilt}) as reported in \citet{Sarkar_2020}. \\

\begin{deluxetable}{ll}[!b]
\tablecaption{Model runs with different density values for the CME \label{table2}}
\tablehead{
\colhead{Run number} & \colhead{Spheromak density} 
}

\startdata
Run1 (LD) & 1$\times10^{-18}$\;kg m$^{-3}$ \\
Run2      & 2$\times10^{-18}$\;kg m$^{-3}$ \\
Run3      & 4$\times10^{-18}$\;kg m$^{-3}$ \\
Run4      & 6$\times10^{-18}$\;kg m$^{-3}$ \\
Run5      & 8$\times10^{-18}$\;kg m$^{-3}$ \\
Run6 (MD) & 10$\times10^{-18}$\;kg m$^{-3}$ \\
Run7 (HD) & 50$\times10^{-18}$\;kg m$^{-3}$ \\
\enddata
\end{deluxetable}

\subsubsection{Radius} The geometrical reconstruction of the 3D morphology of a CME can be approximated using the GCS model. The geometry of the GCS model provides both the edge-on and face-on radii of a CME flux-rope that resembles a croissant-like structure. To transfer this shape to the spherical case, we average the edge-on and face-on radii of the GCS structure to determine the spheromak radius. Notably, we assume self-similar expansion of the CME up to 21.5\;$R_S$ from the last location where the 3D reconstruction was performed. Following the aforementioned assumption, we consider that the edge-on and face-on angular width of the CME as obtained from GCS fitting remains constant up to 21.5\;$R_S$. Using the fitted edge-on and face-on angular width as reported in \citet{Sarkar_2020}, we obtain the radius of the spheromak as 7.2\;$R_S$ if its leading edge locates at 21.5\;$R_S$.

\subsubsection{Speed and propagation direction} The LASCO CME catalogue provides a linear speed of the CME as 861\;km~s$^{-1}$, projected in the plane-of-sky as viewed from Earth. Applying the GCS fitting to the multi-vantage point white-light observations of the CME at different instances of time, we find the de-projected 3D speed of the CME to be $\approx$ 930\;km~s$^{-1}$. Notably, the speed (V$_{3D}$) obtained from the GCS fitting can be decomposed as the translational speed (V$_{radial}$) and the expansion speed (V$_{exp}$) of the structure. The expansion of the linear force-free (LFF) spheromak due to the pressure imbalance with the ambient solar wind is modelled self-consistently with EUHFORIA. Therefore, it is required to only use the V$_{radial}$ as the input speed to run the simulation. We utilize the empirical relationship, $V_{radial} = 0.43\times V_{3D}$, as used in \citet{2019scolini} to obtain the input translational speed of the LFSS as 400\;km~s$^{-1}$.    

Following the GCS results on this event studied by \citet{Sarkar_2020} and \citet{Vemareddy_2015}, we notice that the propagation direction of the CME ranges between 9$^{\circ}$N-15$^{\circ}$S in latitudinal and 0$^{\circ}$-13$^{\circ}$E in longitudinal direction in HEEQ. Throughout the paper, we guide the reader about the spheromak evolution by presenting its cross-sectional map on the HEEQ equatorial plane and take the Sun-earth line as reference line with respect to which we show any change in orientation of the spheromak magnetic axis curve projected on the equatorial plane. Therefore, we select 0$^{\circ}$ longitude and  0$^{\circ}$ latitude in HEEQ (within the above-mentioned observational range) as the insertion direction of the spheromak, as in such scenario the equatorial plane approximately cut through the spheromak centroid, and it propagates nearly along the Sun-Earth line. We further assess the effects of uncertainties related to the direction of propagation by comparing the model results from multiple virtual spacecraft located within 15$^{\circ}$E-15$^{\circ}$W and 15$^{\circ}$N-15$^{\circ}$S at 1~au (see Section~\ref{insitu_assess}).

\subsubsection{Helicity sign}
Observations of the inverse S-shaped morphology of the pre-flare sigmoid structure as reported in \citep{Vemareddy_2015, Sarkar_2020} suggests a left-handed chirality of the associated flux-rope. This helicity sign is also consistent with the hemispheric helicity rule as the source active region of the CME was located in the northern hemisphere. We therefore set the chirality of the spheromak to be negative. 

\subsubsection{Magnetic flux} 
Observational studies show that the poloidal flux of a CME can be obtained from the estimation of reconnection flux as computed from cumulative flare ribbon area \citep{Kazachenko_2017} or post-eruption arcades \citep{2018Gopal}. In order to constrain the magnetic flux of a spheromak 
%at the lower boundary of heliospheric domain 
in EUHFORIA, \citet{2019scolini} equates the observationally estimated poloidal flux to that of a spheromak. However, we notice that the magnetic axis of a spheromak resembles that of a full torus-like structure without having any anchoring points on the Sun. On the other hand, the observed poloidal flux is believed to be distributed over a nearly half-torus like structure with an angular extent similar to the side-on angular width of a CME. Therefore, equating the observed poloidal flux to that of a spheromak would underestimate the field-strength at its magnetic axis due to the distribution of observed poloidal flux over a comparatively large volume of a spheromak.

In order to avoid the above-mentioned underestimation, we follow a different approach to constrain the magnetic flux of the spheromak. Instead of directly equating the observed poloidal flux with that of the spheromak, we equate the field strength ($B_\mathrm{spheromak}$) at the magnetic axis of the spheromak to the axial field strength ($B_0$) obtained from the Flux Rope Eruption Data (FRED) technique \citet{IAU_Gopalswamy}. Conceptually, FRED uses the information on poloidal flux and a Lundquist flux-rope model \citep{lundquist} constrained with realistic CME geometry to estimate the axial field strength of a CME. Further, using the value of $B_\mathrm{spheromak}$, we estimate its toroidal flux content which is used as input for the EUHFORIA simulation. Therefore, the axial field strength ($B_\mathrm{spheromak}$) of a spheromak remains identical to that obtained from FRED and does not depend on the spheromak volume.   

\citet{Sarkar_2020} reported the average observed poloidal flux of the associated CME under study as 2.1$\times$10$^{13}$ Wb and the axial field strength ($B_0$) at 10\;\RSun\ as 5200\;nT. Applying the FRED technique further out to the inner boundary (21.5\;\RSun) of EUHFORIA, we obtain $B_0$ at 14.3~$\RSun$ (21.5 $\RSun$ - r$_{sph}$) as 2500~nT. Notably, each point on the magnetic axis of a spheromak is not equidistant to the Sun-center due to the circular shape of the axis. Therefore, we use the radial distance (14.3~$\RSun$) of the spheromak center (21.5~$\RSun$ - r$_{sph}$) for deriving $B_0$ which is the average distance of all the points on its magnetic axis. Equating the estimated $B_0$ value (2500\;nT) to $B_\mathrm{spheromak}$, we determine the toroidal flux of the spheromak as 2.4$\times$10$^{13}$ Wb and use this value as input for EUHFORIA simulation. 

\begin{figure*}[t!]
\begin{center}
\includegraphics[width=\textwidth,clip=]{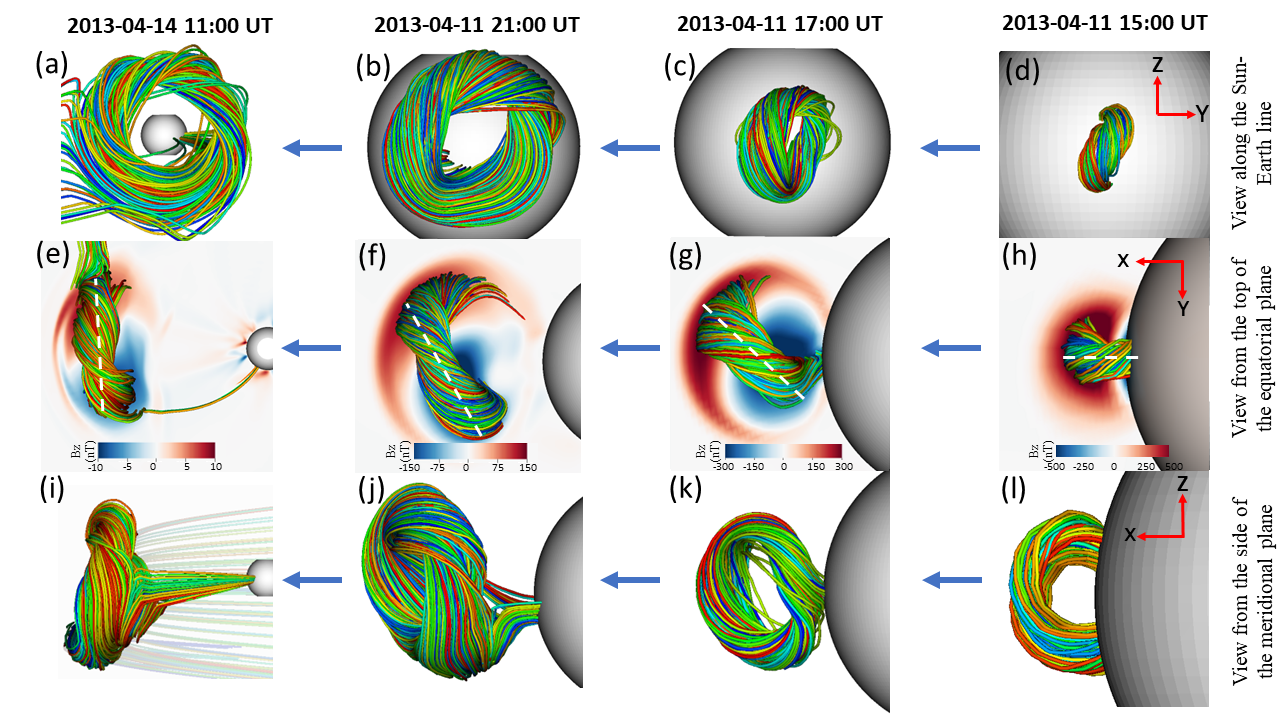}
\caption{Rotation of the spheromak during different phases of its interplanetary evolution from 0.1~au to 1.0~au as visualised from three different viewpoints. The view along the Sun-Earth line (a, b, c \& d), from the top of the equatorial plane (e, f, g \& h) and from the side of the meridional plane (i, j, k \& l) are shown in top, middle and bottom rows respectively. The temporal sequences advance in time from right to left as shown by the blue arrows. The right-most column showcases the spheromak during its insertion at 0.1~au (d, h \& l). The other columns from right to left display the spheromak evolution when its front approximately reaches at 0.24~au (c, g \& k), 0.28~au (b, f \& j) and 1.0~au (a, e \& i) respectively.
The bunch of twisted lines represent the magnetic field lines wrapping around the spheromak magnetic axis. The grey sphere is the lower boundary of the simulation domain at 0.1~au. The red and blue colours in the middle row indicate the direction (northward and southward respectively) of $B_z$ on the equatorial plane. The white-dashed line drawn over the selection of field lines in the middle row indicates the orientation of the magnetic axis in the equatorial plane.\label{spheromak_rotation}}
\end{center}
\end{figure*}

\begin{figure*}[t!]
\begin{center}
\includegraphics[width=.9\textwidth,clip=]{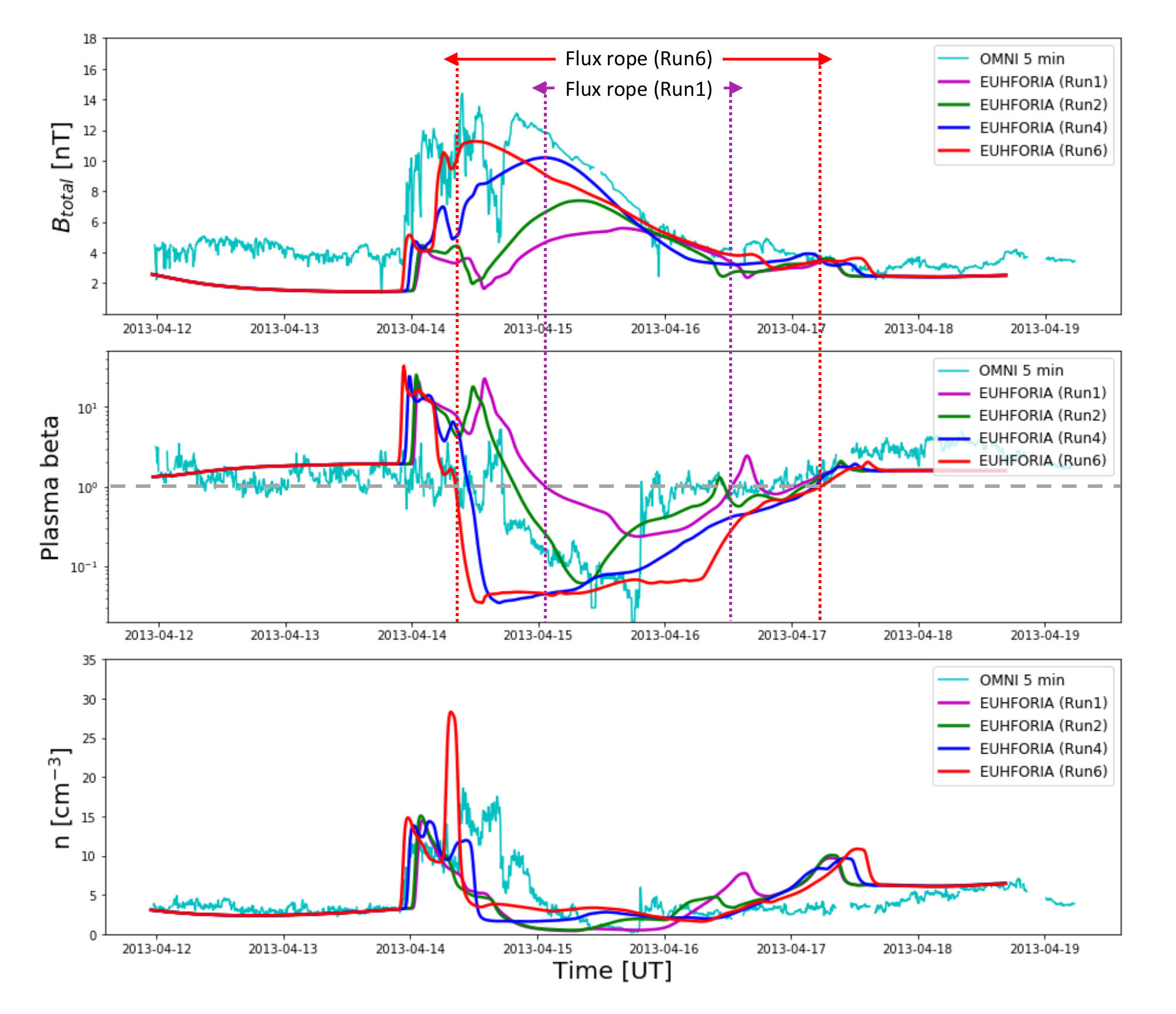}
\caption{The magnetic field strength (top panel), plasma beta (middle) and the density (bottom panel) of the ICME obtained from the model runs using different spheromak densities plotted on top of the observed in-situ values at 1~au. The grey dashed horizontal line in the middle panel indicates a plasma beta value of one. The vertical dotted lines in red and magenta mark the flux-rope boundaries for Run6 and Run1, respectively, which are identified based on the plasma beta values that are less than one.}
\end{center}
\label{density_mag}
\end{figure*}%\label{density_mag}

\begin{figure*}[t!]
\begin{center}
\includegraphics[width=\textwidth,clip=]{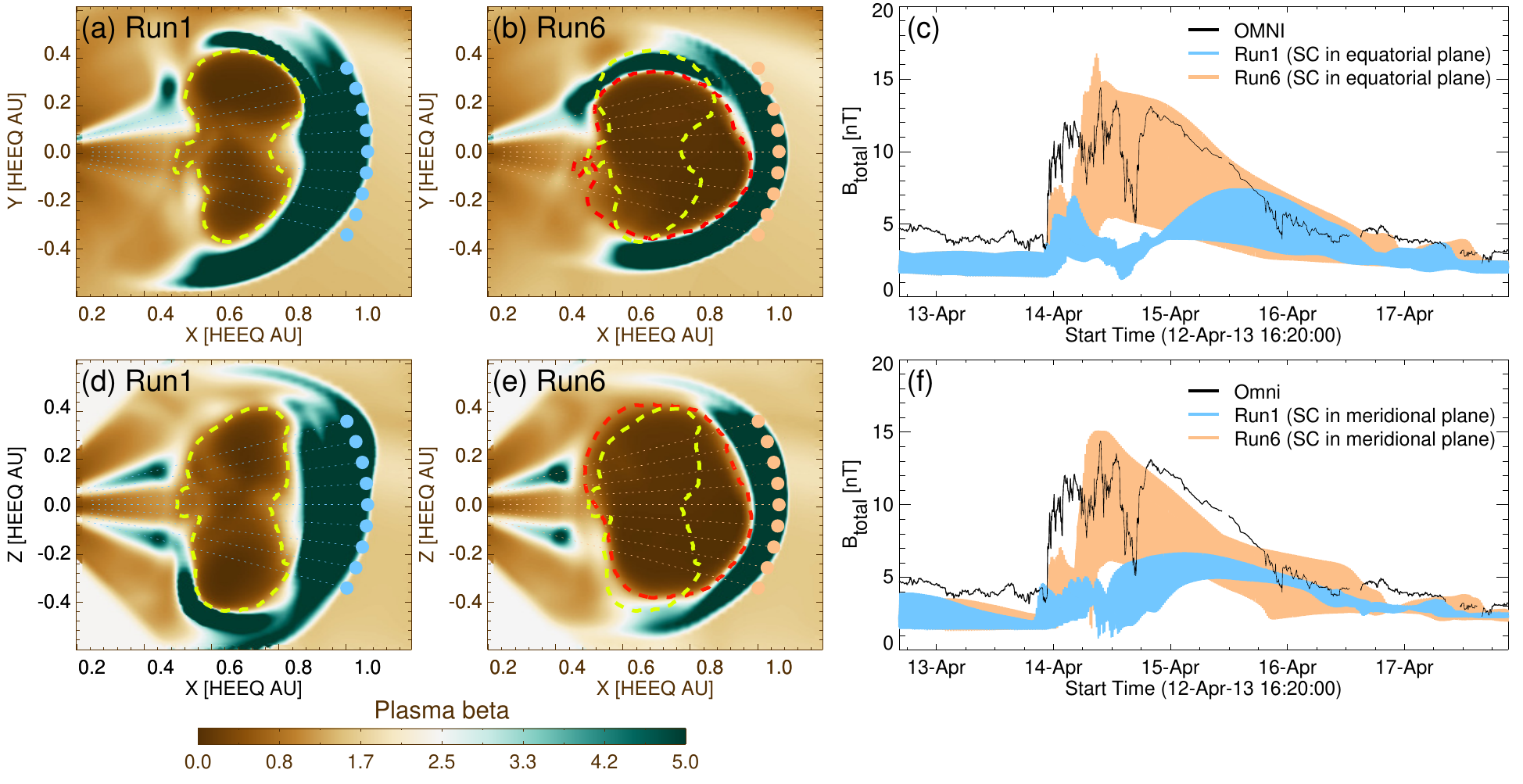}
\caption{Plasma beta in the equatorial plane for the low [Run1] (a) and moderate [Run6] (b) density runs are plotted when the shock front ahead of the spheromak approximately reaches 1~au. The yellow/red dashed contour in panel (a)/(b) encloses the region where the plasma beta value is less than 1 within the ICME structure. The same yellow dashed contour shown in panel (a) is over-plotted in panel (b) to contrast the size of the spheromaks in the two different runs. Panel (c) displays a comparison of the observed magnetic field strengths of the ICME and the model results for the various virtual spacecraft (SC) in the ecliptic plane located within $\pm$ 20$^{\circ}$ with respect to Earth at 1~au. Panel (d), (e) and (f) are same as panels (a), (b) and (c) respectively, but represent the results in the meridional plane.}
\end{center}
\label{sc_mag}
\end{figure*}

\subsubsection{Temperature \& density:}
The default values for the CME mass density (1$\times10^{-18}$\;kg m$^{-3}$) and temperature (0.8$\times10^{6}$\;K) are widely used in studies employing EUHFORIA and are set to be uniform inside the CMEs \citep{2018Pomoell,2019scolini, Asvestari1}. For the initial simulation run (Run1), we use these default values of the CME mass density and temperature as input for the LFFS. 

We further study the dependence of the CME evolution on the choice of the initial mass density by conducting a set of EUHFORIA simulations with density values higher than the default one. Five simulations (Run2, 3, 4, 5 \& 6) are performed by selecting the density values within the observational range (1$\times10^{-18}$\;kg m$^{-3}$ to 1$\times10^{-17}$\;kg m$^{-3}$) at 0.1~au \citep{2021Mannuela} in steps of 2$\times10^{-18}$\;kg m$^{-3}$ (see Table~\ref{table2}). We discuss the results based on these runs in Section~\ref{higher_densities}. We notice that the spheromak centroid height locates at $\approx$ 14\;\RSun\ when we start to insert it at the lower boundary (0.1~au) of the simulation. The observational upper limit of the CME density at $\approx$ 14\;\RSun\ turns out to be $\approx$ 5$\times10^{-17}$\;kg m$^{-3}$ as
reported in \citet{2021Mannuela}. Considering this higher density value, we perform a further simulation (Run 7) and refer to it as the ``high-density (HD) run". In Section~\ref{higher_densities}, we compare the results of the ``high-density run" with Run~6 and Run~1, which are hereinafter referred to as the ``moderate-density (MD) run" and ``low-density (LD) run", respectively. For all runs, we keep the rest of the input parameters unchanged.

For all the runs in this work, a uniform mesh-grid is used with a 2$^{\circ}$ angular resolution and a radial grid spacing of $\Delta r \approx 0.0037$~au $\approx 0.8$ solar radii  (512 cells in radial direction).

\begin{figure*}[t!]
\begin{center}
\includegraphics[width=\textwidth,clip=]{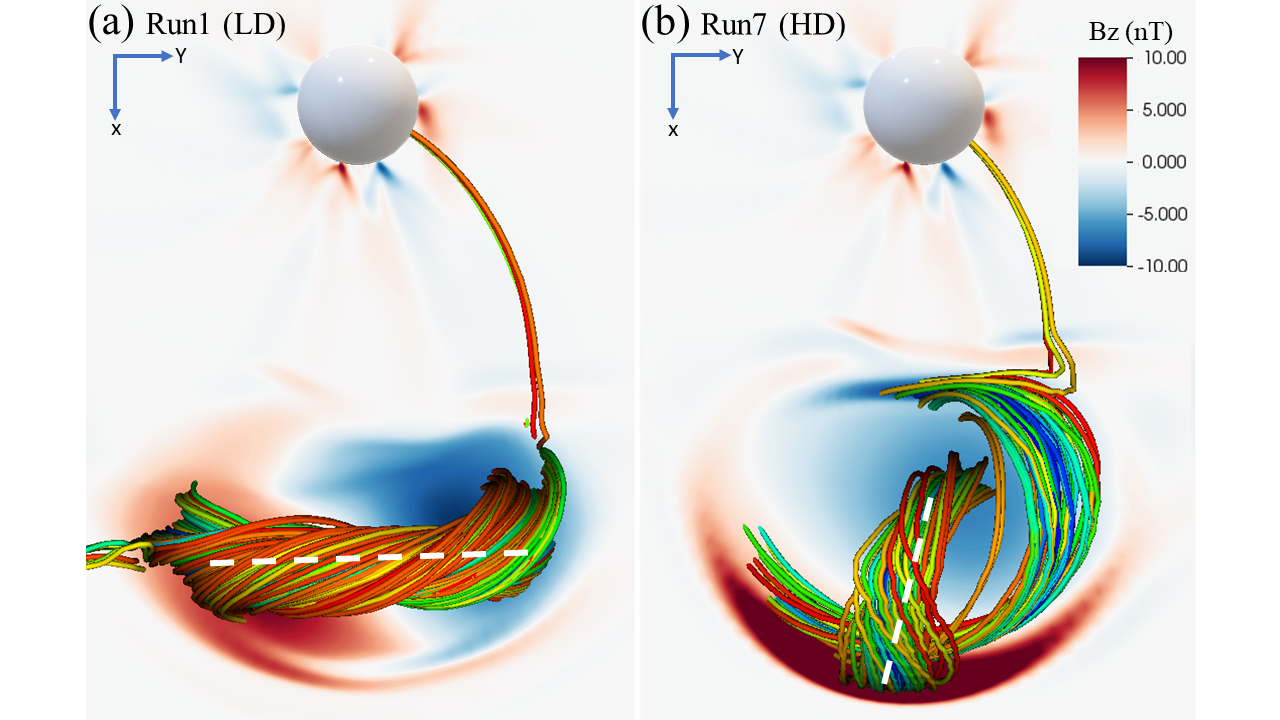}
\caption{Magnetic field configuration of the modeled ejecta with initially low plasma density (Run1; panel a) and high density (Run7; panel b) as viewed from above the equatorial plane. The white-dashed line drawn over the selection of field lines indicates the projected orientation of the magnetic axis curve in the equatorial plane. The white sphere is the lower boundary of the simulation domain at 0.1~au. The red/blue color in the background indicates the magnitude and direction (northward/southward) of $B_z$ in the equatorial plane.}\label{rotation_3d}
\end{center}
\end{figure*}

\section{Results}\label{section4}

\subsection{Spheromak rotation}\label{spheromak rotation}

The evolution of the spheromak in the heliospheric domain as obtained from the simulation results for model Run1 is illustrated in Figure~\ref{spheromak_rotation}. In order to visualise the 3D orientation of its magnetic axis, we plot a selection of field lines that wrap around the magnetic axis of the spheromak. We do not plot the less-twisted magnetic field lines passing close to the symmetry axis of the spheromak. Therefore, in the visualisation of Figure~\ref{spheromak_rotation}, a void structure appears at the central part of the spheromak, shaping it to a torus-like structure (e.g. see panel a or l). Tracking this structure at different instances of time during its propagation from Sun to Earth, the image shows that the axis orientation of the spheromak changes considerably from that of its insertion at 0.1~au.

In the course of the insertion, as depicted in the right-most column (panels d, h \& l) of Figure~\ref{spheromak_rotation}, the spheromak orients in such a way that the central void-part of its analogical torus-like shape is fully observable from the side view on the meridional plane (see panel l) and its magnetic axis curve as projected on the equatorial plane nearly aligns to the x-direction, i.e., the Sun-Earth line (see panel h). As the spheromak propagates further out in the heliosphere, it rotates in such a way that the central void-part of the torus-like shape starts to disappear from the side view of the meridional plane (see the evolution from right to left in lower row of Figure~\ref{spheromak_rotation}, i.e. panels l, k, j and i respectively) and becomes partial to fully visible along the Sun-Earth line (upper row of Figure~\ref{spheromak_rotation}, i.e. panels d, c, b \& a respectively). The evolution of its projected axis on the equatorial plane as shown in the middle row of Figure~\ref{spheromak_rotation}, clearly depicts that it exhibits a clockwise rotation (as viewed from the top of the equatorial plane) of $\approx 90^{\circ}$ in the interplanetary space and becomes almost perpendicular to the Sun-Earth line (x-direction) when it arrives at 1~au. We present a more quantitative analysis of the changes in the orientation angle of the spheromak with increasing heliocentric distances from 0.1~au to 1~au in section~\ref{higher_densities}. 

The significant rotation of the spheromak as obtained from  Run1 is consistent with the finding of spheromak rotation as reported in \citep{Asvestari2}. Indeed, the symmetry axis of the spheromak rotates in such a way that it aligns  approximately along the Sun-Earth line. As a result, the virtual spacecraft at Earth encounters the least twisted part of the spheromak along its symmetry axis and misses the part of the magnetic structure that is constrained from observations (see Figure~\ref{ini_tilt}). This can be a significant issue for space weather modelling when using spheromak as CMEs. 

Notably, we performed Run1 with the default density value for spheromak which has been used  in previous studies with EUHFORIA. We further present the results of a set of simulations (see Table~\ref{table2}) performed with higher density values of spheromak in Section~\ref{higher_densities}.

\subsection{Spheromak evolution with higher densities}\label{higher_densities}
We find that the spheromaks with higher densities result in higher magnetic-field strengths at the position of Earth. We present this result in Figure~\ref{density_mag}, which shows the temporal profiles of spheromak field strength (top panel) at 1~au from a selection of runs performed with different densities. Notably, a spheromak with higher density is expected to undergo higher expansion due to the increased internal thermal pressure. Indeed, the comparison of the model results at the position of Earth as obtained from Run1 and Run6 (Figure~\ref{density_mag}), depicts that the spheromaks with initial higher density become larger in size (enclosed by the two vertical dotted lines in red in the figure) as compared to the ones with initially lower density (enclosed by the two vertical dotted lines in magenta) based on their appearance in the in-situ observations by a virtual spacecraft at Earth. Therefore, due to the higher internal expansion, one would expect that the high density spheromaks will result in a correspondingly lower magnetic field strength upon its arrival at 1~au, which is in contrast to our results as mentioned at the beginning of this section (top panel of Figure~\ref{density_mag}). Notably, different trajectories of the spacecraft through the spheromak can significantly contribute to the changes in its magnetic strength profile and the inferred size. If the amount by which the spheromak rotation varies due to varying the initial mass density, then the virtual spacecraft at Earth would probe different parts of the spheromak for different cases. Therefore, we further compare the results of Run1 and Run6 for several virtual spacecraft placed at different HEEQ longitudes and latitudes within $\pm$ 20$^{\circ}$ from the nominal position of Earth at 1~au (see Figure~\ref{sc_mag}). The plasma beta of the spheromak for Run1 and Run6 in both equatorial (Figure~\ref{sc_mag}[a] \& [b]) and the meridional plane (Figure~\ref{sc_mag}[d] \& [e]), clearly show that the size of the spheromak is larger for Run6 (indicated by the red dashed contour) as compared to Run1 (indicated by the yellow dashed contour), which is in agreement with our results based on Figure~\ref{density_mag}. Figure~\ref{sc_mag}[c] \& [f] shows that the magnetic field strength profiles obtained at the different virtual spacecraft for Run6 (orange shaded profile) are predominantly higher as compared to that for Run1 (cyan shaded profile). This again indicates that the high density spheromaks lead to higher field strengths at 1~au as also seen in Figure~\ref{density_mag}. However, the considerable variability in magnetic field strength observed as virtual spacecraft traverse distinct regions within the spheromak for both Run1 and Run6 (Figure~\ref{sc_mag}[c] \& [f]), highlights the notion that if the spheromak undergoes varying degrees of rotation due to different densities, it would result in diverse magnetic profiles observed at Earth. On the other hand, the predominantly higher field strength obtained for the higher density run (Run6) incorporating the different spacecraft crossing through the spheromak, suggests that the variance in spheromak rotation alone cannot account for these observations, implying the existence of an additional compressional influence acting upon the spheromak. Therefore, we suggest the following two possible scenarios that contribute to the diverse spheromak profiles observed at Earth under varying density conditions:   

(i) The spheromaks rotate differently when their density increases. Therefore, the virtual spacecraft located at the position of Earth encounter different parts of the spheromak for the runs performed with higher density values which results in different field strength at 1~au.

(ii) The spheromaks with higher density undergo compression which leads to higher field strength at 1~au when compared to the low-density cases.

We explore the above-mentioned two possibilities based on our results in the following sub-sections~\ref{section_rotation} and \ref{section_expansion} respectively.

\begin{figure}[t!]
\begin{center}
\includegraphics[width=.5\textwidth,clip=]{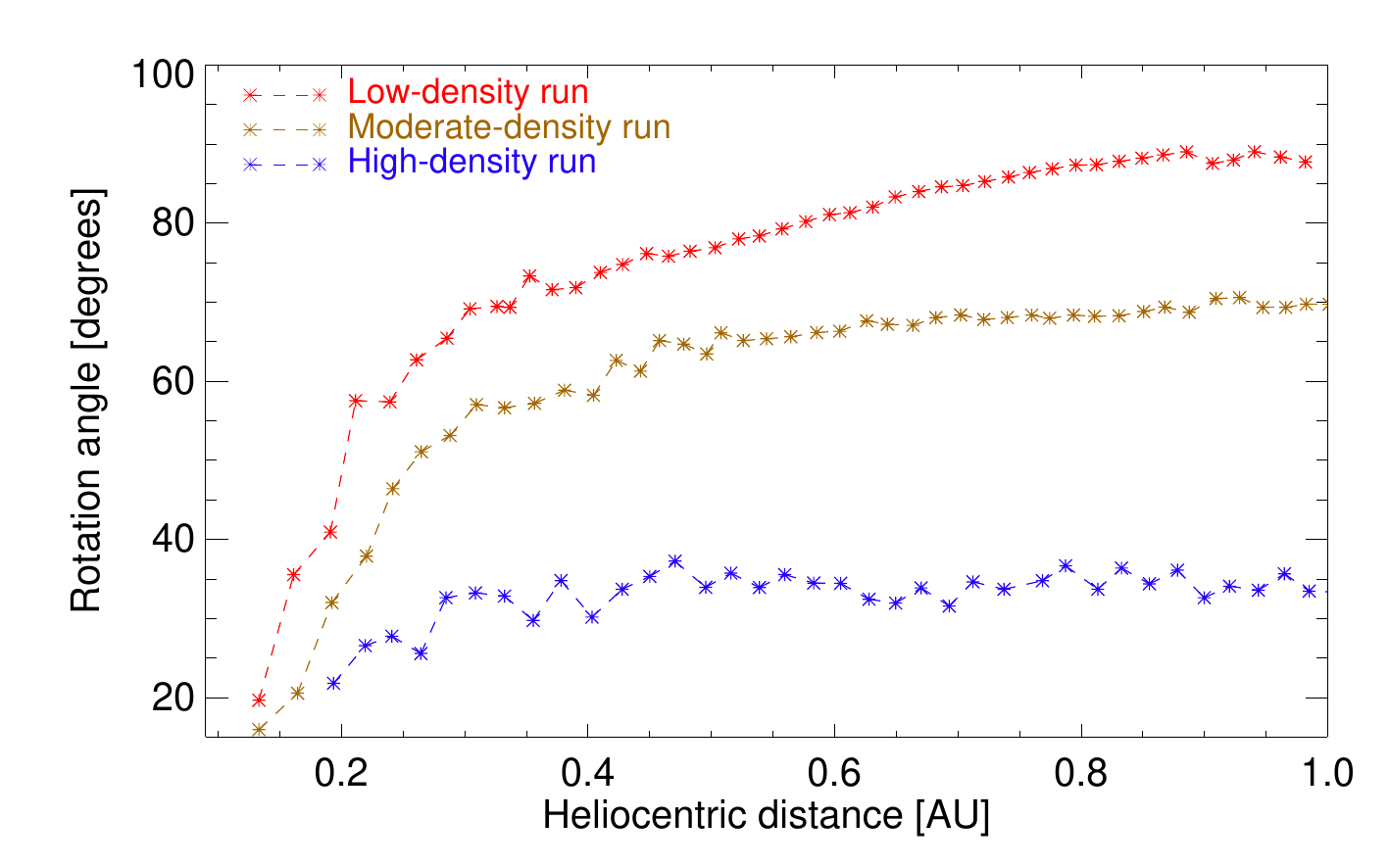}
\caption{Change in rotation angle of the spheromak magnetic axis with increasing heliocentric distance from Sun as obtained from the different density runs.}\label{rotation_comparison}
\end{center}
\end{figure}

\begin{figure*}[t!]
\begin{center}
\includegraphics[width=.9\textwidth,clip=]{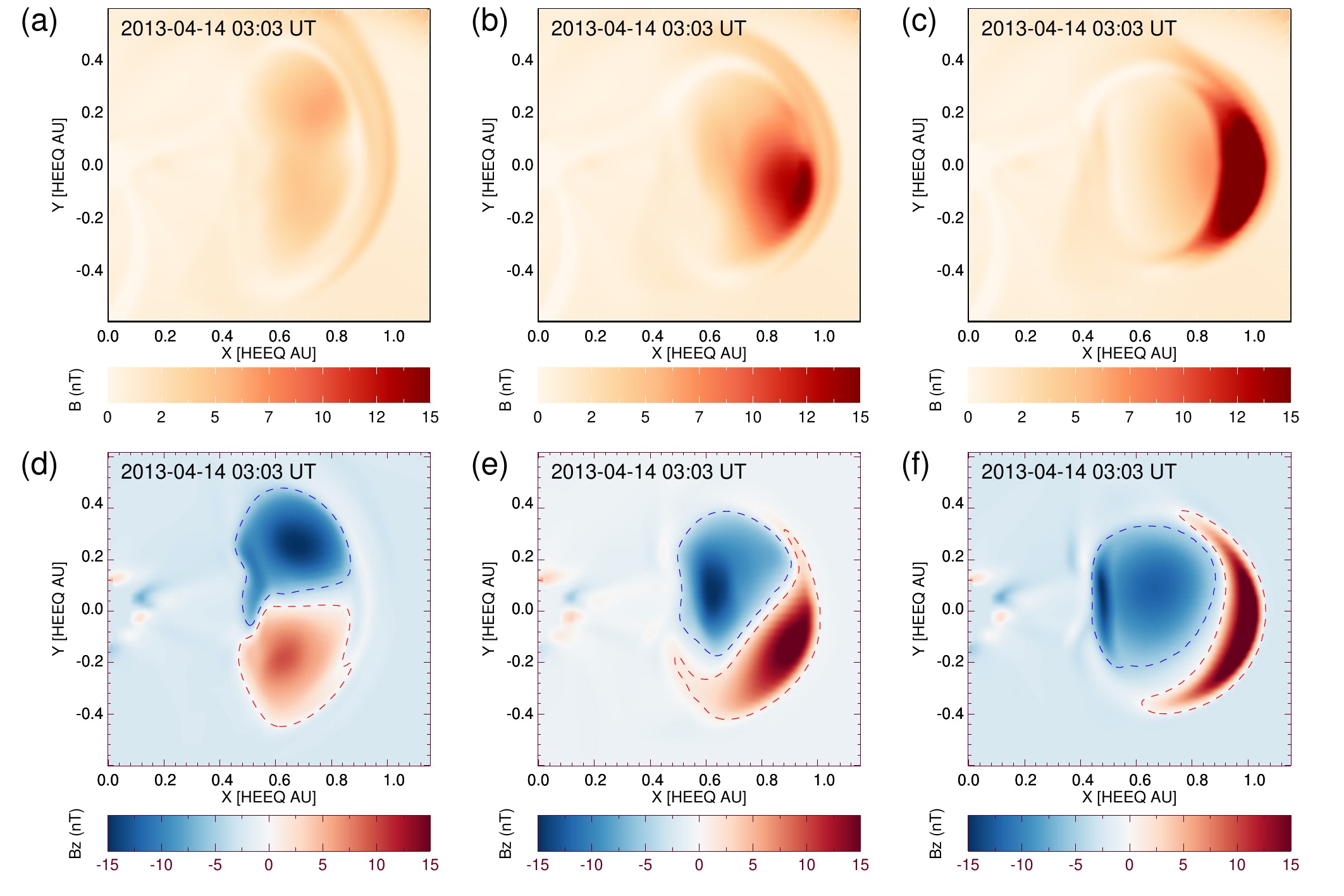}
\caption{Magnetic field strength of the spheromak in the equatorial plane for the low [Run1] (a), moderate [Run6] (b) and high [Run7] (c) density runs. The $B_z$ component of the spheromak in HEEQ for the low [Run1] (d), moderate [Run6] (e) and high [Run7] (f) density runs.}\label{density_compression}
\end{center}
\end{figure*}

\subsubsection{Effect of density on spheromak rotation}\label{section_rotation}
The 3D configuration of the magnetic field of the CME around the magnetic axis at the time when the leading edge of the CME reaches 1~au for the low (panel~a) and high-density (panel~b) runs is visualised in Figure~\ref{rotation_3d}. The white-dashed lines over-plotted on each panel of Figure~\ref{rotation_3d} depict the approximate projection of the magnetic axis curve on the equatorial plane. The image clearly illustrates that at 1~au the orientation of the magnetic axis for the high-density run is significantly different from that for the low-density run. Interestingly, we find that the magnetic axis associated with the high-density run undergoes less rotation, as the projected orientation of the magnetic axis curve at 1~au is nearly aligned with its initial orientation (approximately along the Sun-Earth line; see panel~h of Figure~\ref{spheromak_rotation} and panel~b of Figure~\ref{rotation_3d} for comparison) at the time of insertion. 

We further present a quantitative analysis of the change in the axis orientation of the modeled CME as it evolves from 0.1~au to 1.0~au launched with different densities. A detailed description of the methodology to estimate its orientation angle is given in the Appendix. Figure~\ref{rotation_comparison} depicts the evolution of the magnetic axis orientation (projected on the equatorial plane) with increasing heliocentric distances for the high, moderate and low-density runs, respectively. Note that the orientation angle is measured with respect to the Sun-Earth line which is the approximate direction of propagation for this  event. It is clearly delineated in Figure~\ref{rotation_comparison} that while the magnetic axis undergoes a large rotation ($\approx 90^{\circ}$) for the low-density run, the amount of rotation is significantly lower for the moderate  ($\approx 70^{\circ}$) and in particular so for the high density ($\approx 35^{\circ}$) runs before it arrives at 1~au. Importantly, a significant part of the rotation for all the runs occurs below 0.3~au. This indicates the role of the strong ambient magnetic field that exerts higher torque on the ejected structure when it is close to the Sun \citep{Asvestari2}. Our finding of the density dependence on the change of orientation suggests that the ambient magnetic field becomes less effective in rotating the more massive spheromaks due to the higher linear momentum of the spheromak towards its direction of propagation. Therefore, the high-density spheromaks exhibit less rotation than the low-density ones.

\subsubsection{Effect of density on the internal expansion of the spheromak}\label{section_expansion}
The choice of the initial plasma density in the spheromak does not only change the amount of rotation it experiences during its interplanetary propagation but affects its internal expansion as well. The magnetic field magnitude associated with the spheromak and the north-south component ($B_z$ in HEEQ) of the magnetic field in the HEEQ equatorial plane are shown in Figure~\ref{density_compression}. The $B_z$ component for low, moderate and high-density runs as shown in the bottom panels of Figure~\ref{density_compression}, shows two separate regions of oppositely directed $B_z$ that are made more evident by the enclosed red and blue dashed contours drawn at levels $\pm$ 2.5\;nT respectively. For the sake of further explanation, we refer to these two regions of dominant positive and negative $B_z$ as red- and blue-regions following the color-map as depicted in the figure. In particular, these red and blue-regions as shown on the cross-sectional slices of the spheromak contain the twisted field lines (see Figure~\ref{rotation_comparison}) centering its magnetic axis which is pointing in and out of the equatorial plane respectively.  

Notably, the magnetic axis orientation of the spheromak in the red-region bearing the positive $B_z$ in Figure~\ref{density_compression}, points towards the northward direction which has been constrained from the observed axial direction (see Figure~\ref{ini_tilt}) of the CME. Being a closed curve (see Figures~\ref{spheromak_rotation}~[a] and \ref{spheromak_rotation}~[b]), the magnetic axis of the spheromak turns its direction from the red-region to the blue-region and becomes southward in the blue-region. During the insertion phase of the spheromak, the red-region enters into the simulation first, followed by the blue region (see Figures~\ref{spheromak_rotation} [h] and \ref{spheromak_rotation} [g] sequentially) as the spheromak itself is an isolated magnetic structure and does not have any legs attached to the Sun. However, instead of propagating one-followed-by-another, the two aforementioned regions of the spheromak for the low-density run, approximately move parallel to each other (see Figure~\ref{density_compression}[d]) along the direction of propagation due to its significant rotation below 0.3~au. Therefore, during the majority of the propagation path (from approximately 0.3~au onward), both regions remain in contact with the solar wind ahead. This causes similar interplanetary evolution of both regions which results in their similar shapes at 1~au (see Figure~\ref{density_compression}[d]).

\begin{figure*}[t!]
\begin{center}
\includegraphics[width=.42\textwidth,clip=]{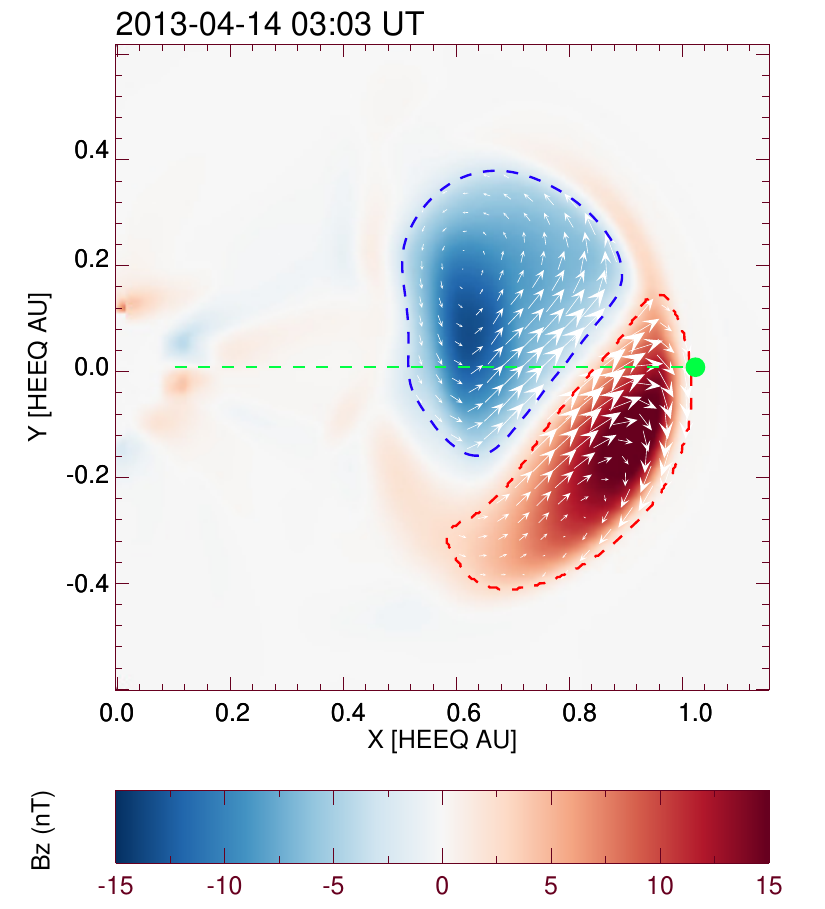}
\includegraphics[width=.54\textwidth,clip=]{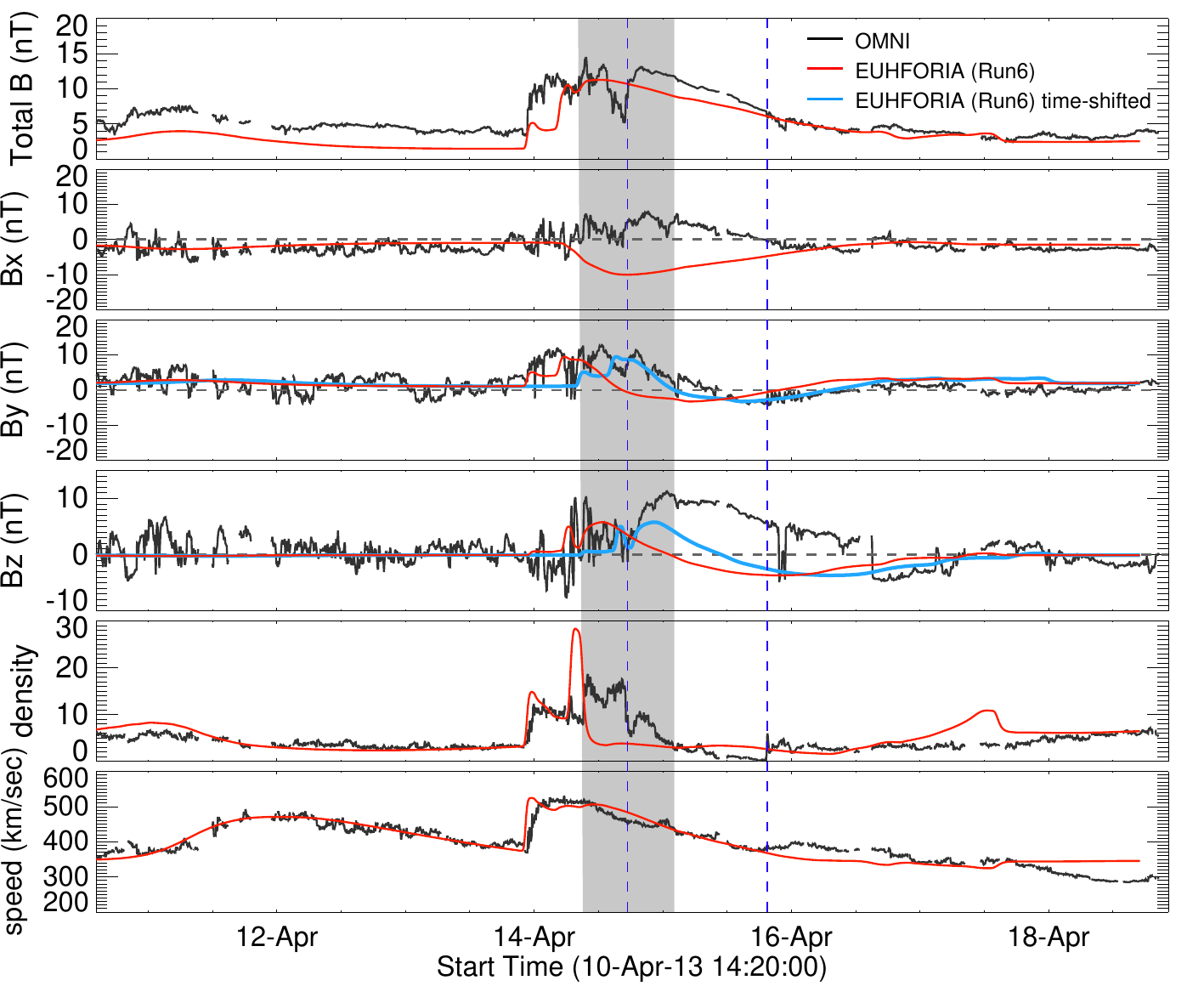}
\caption{The $B_z$ component of the spheromak associated with the moderate density run [Run6] is shown on the equatorial plane (left panel). The blue/red dashed contours enclose the regions with negative/positive $B_z$ inside the spheromak. The white arrows within the blue/red dashed contoured regions denote the magnetic vectors parallel to the equatorial plane. The green dashed line is the Sun-Earth line and the green dot depicts the location of Earth. Comparison of the model results with the observed in-situ plasma properties at 1~au (right panel). The black solid curves denote the observation from WIND and the red solid curves are the model outputs at 1~au. The vertical blue dashed lines mark the boundary of the observed magnetic cloud. The grey-shaded region in the right panel is the temporal passage of the positive $B_z$ portion inside the spheromak as indicated by the region enclosed by the red dashed contour in the left panel. The cyan solid lines in the $B_y$ and $B_z$ plots in the right panel are the same as the model output as indicated by the red solid lines but shifted by 9 hours and 40 minutes so that the front edge of both observed and modelled flux-rope temporally coincides. The dark grey dashed horizontal lines in the $B_x$, $B_y$, and $B_z$ plots indicate a zero value for the respective variables.}\label{in_situ_compare1}
\end{center}
\end{figure*}

However, for the high-density run in absence of any significant rotation, the regions of opposite $B_z$ polarity move approximately one after another along the propagation direction (see Figure~\ref{density_compression}[f]). Therefore, in this case, the northward and southward field portions in the spheromak interact differently with the background solar wind as the negative $B_z$ portion no longer remains in contact with the solar wind ahead. The drag force due to the upstream solar wind ahead acts almost entirely on the positive $B_z$ portion (enclosed by the red dashed contour) at the front which results in a curved and compressed structure of the frontal positive $B_z$ part (see Figure~\ref{density_compression}[f]). Due to this compression, the field strength at the frontal part of the spheromak enhances as depicted by the color map in Figure~\ref{density_compression}[c] and [f]. More precisely, the maximum field strength of the spheromak as it passes through the virtual spacecraft at 1~au for this case (Run7) reaches 19.3\;nT which is almost two times larger than that (11.3\;nT) as obtained in Run6 (see the top panel of Figure~\ref{density_mag}). On the other hand, the rear-ward negative $B_z$ portion of the spheromak gets shielded from the upstream solar wind by the frontal portion and therefore does not undergo any significant compression or change in shape.    

An intermediate situation arises for the moderate-density run (Figure~\ref{density_compression}[b] and \ref{density_compression}[e]) where the magnetic field structure undergoes less rotation as compared to the low-density run but experiences a larger rotation as compared to the high-density run. In this scenario, the region bearing the positive $B_z$ value partially shields the other region located at a smaller heliocentric distance and experiences less compressive interaction with the upstream solar wind as compared to the high-density run. Therefore, the results presented in this section clearly explain why the high-density spheromaks result in higher magnetic field strength at 1~au as depicted in Figure~\ref{density_mag}.

\begin{figure*}[t!]
\begin{center}
\includegraphics[width=\textwidth,clip=]{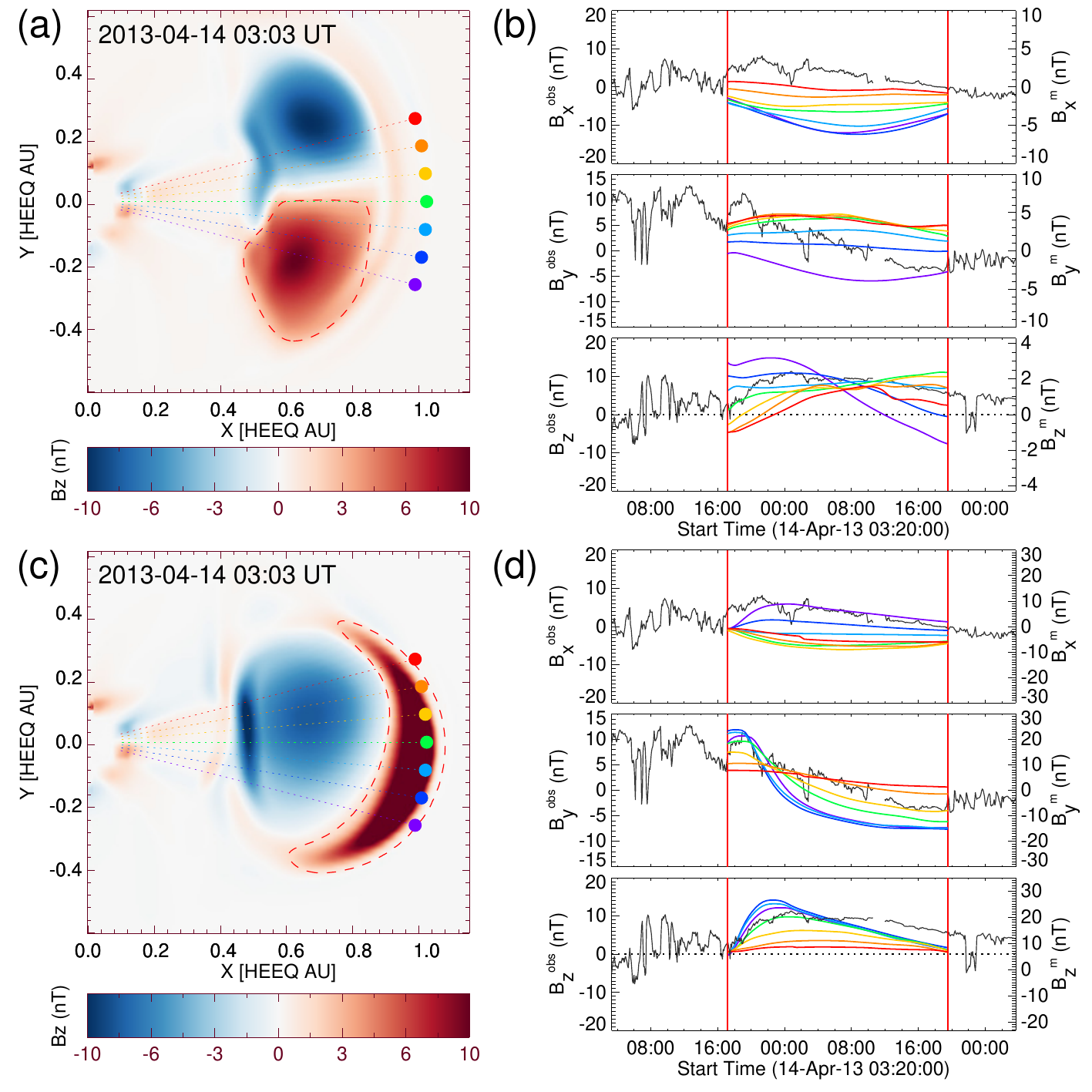}
\caption{The HEEQ $B_z$ component of the spheromak associated with the low (panel [a]) and high (panel [c]) density run are shown on the equatorial plane. The coloured dots drawn on the equatorial plane are the locations of the virtual spacecraft placed at 1~au. Each consecutive spacecraft are longitudinally separated by 5$^{\circ}$, where the red/purple dot corresponds to +/- 15$^{\circ}$ longitude in HEEQ and the central dot in green denotes the location of Earth. The red dashed contour encloses the domain with positive $B_z$ inside the spheromak. The temporal profiles of magnetic field vectors inside the spheromak as it passes through the respective virtual spacecraft  for low and high density runs are shown in panels [b] and [d]. The above mentioned modelled magnetic field profiles (${B_x}^m$, ${B_y}^m$ and ${B_z}^m$) are plotted with a color that is same as that of the associated virtual spacecraft shown in the panels [a] and [c]. The black solid curves are the observed in-situ magnetic field (${B_x}^{obs}$, ${B_y}^{obs}$ and ${B_z}^{obs}$). In panel [b], the modelled magnetic profiles obtained at the virtual spacecraft during the whole spheromak crossing are plotted within the observed MC boundaries indicated by the red vertical lines. In panel [d], only the modelled magnetic profiles obtained at the virtual spacecraft during the passage of the positive $B_z$ domain (the region enclosed by the red dashed contour in panel [c]) of the spheromak are plotted within the observed MC boundaries to avoid the double flux-signature inside the spheromak while assessing the model results with observations.}\label{unc_1}
\end{center}
\end{figure*}

\begin{figure*}[t!]
\begin{center}
\includegraphics[width=\textwidth,clip=]{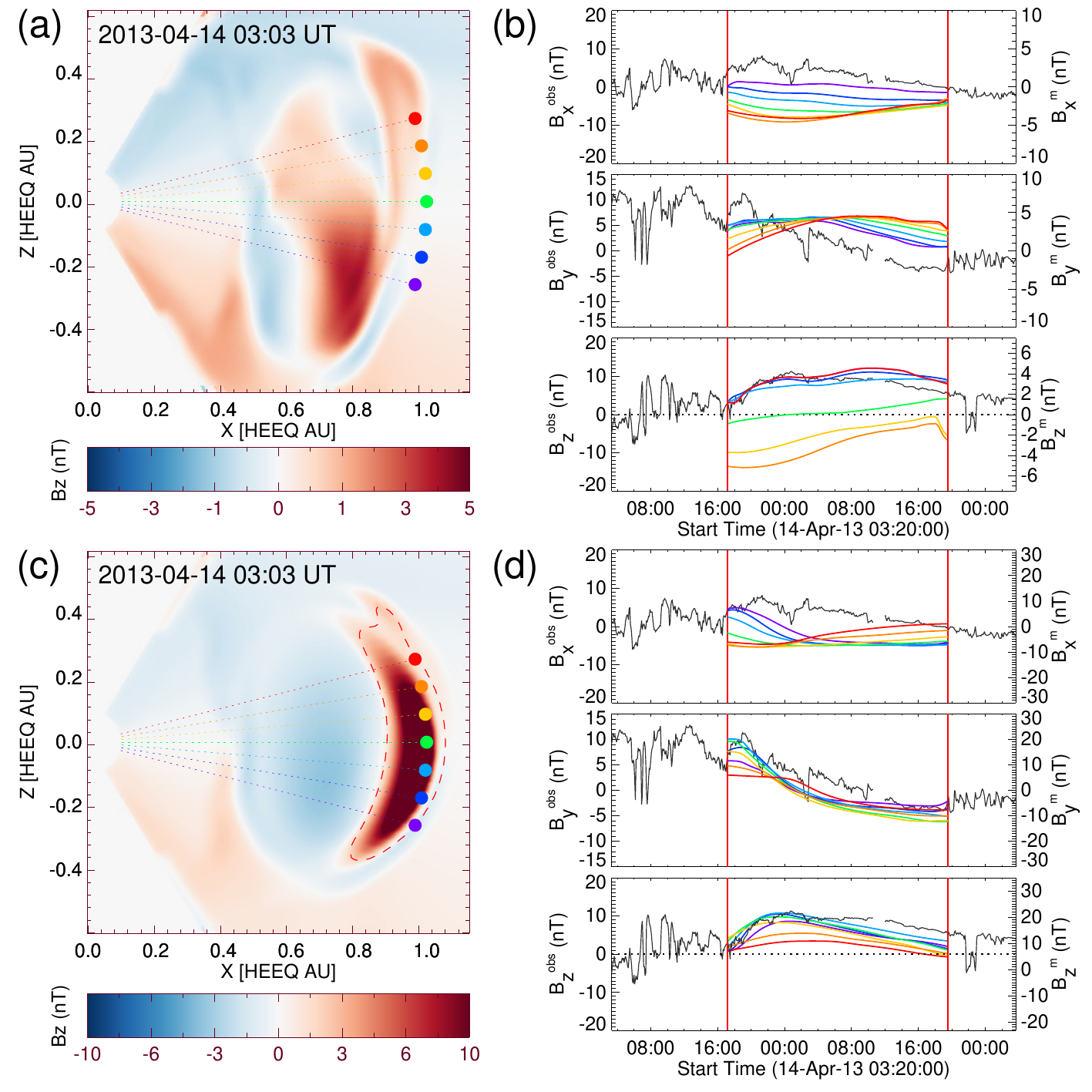}
\caption{The $B_z$ component of the spheromak associated with the low (panel [a]) and high (panel [c]) density run are shown on the meridional plane. The coloured dots drawn on the meridional plane are the locations of the virtual spacecraft placed at 1~au. Each consecutive spacecraft are latitude-wise separated by 5$^{\circ}$, where the red/purple dot corresponds to +/- 15$^{\circ}$ latitude in HEEQ and the central dot in green denotes the location of Earth. The rest of the descriptions of this figure are same as that of the Figure~\ref{unc_1}.}\label{unc_2}
\end{center}
\end{figure*}

\begin{figure}[t!]
\begin{center}
\includegraphics[width=.48\textwidth,clip=]{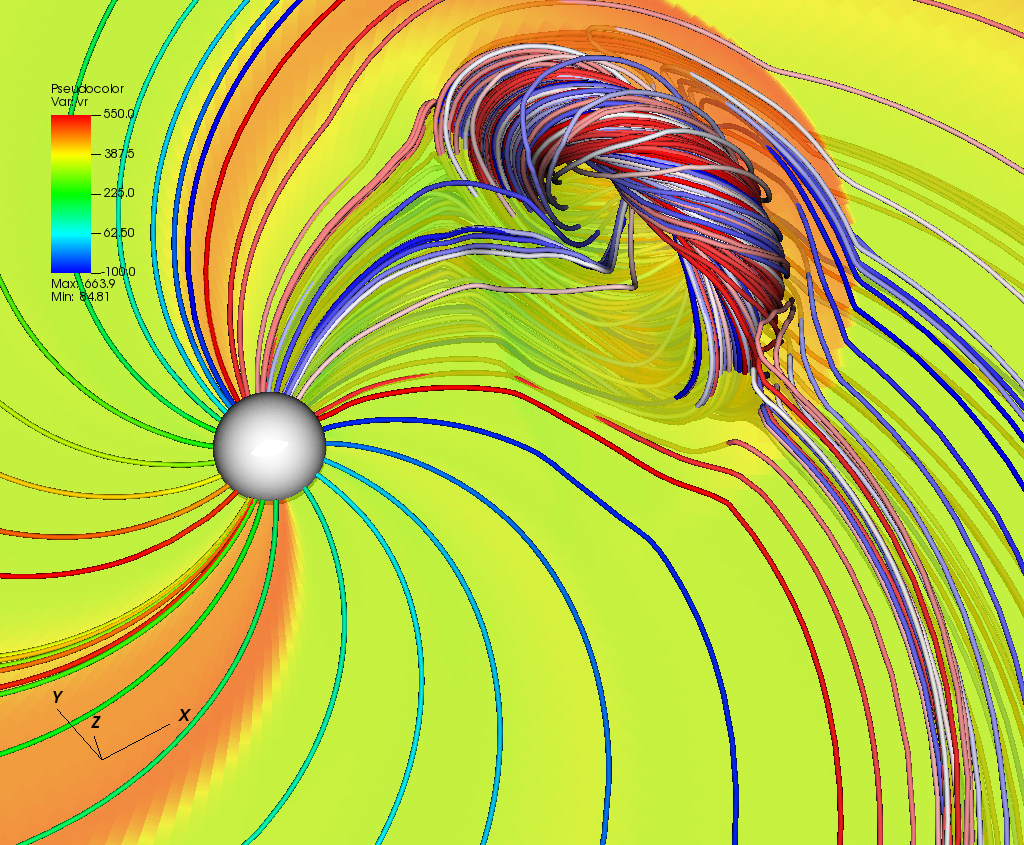}
\caption{Visualisation of the spheromak magnetic structure during its evolution in the heliosphere. The grey-shaded sphere in the image represents the lower boundary of the simulation domain at 0.1 astronomical unit. The dynamic heliosphere is depicted by the background colours of the image which represent the solar wind speed on the equatorial plane. The bunch of twisted lines illustrate the three-dimensional magnetic flux-rope structure of the spheromak propagating away from the Sun. The untwisted field lines emanating from the lower boundary depict the magnetic field topology of the Parker spiral.\label{sheath}}
\end{center}
\end{figure}

\subsection{In-situ comparison}\label{insitu_assess}
The comparison of the in-situ virtual spacecraft observations at Earth with the simulation results in Figure~\ref{density_mag} indicates that the observed magnetic field strength of the ICME is most closely in resemblance to the results of the moderate-density run (Run6). Notably, the input  density (1$\times10^{-17}$\;kg m$^{-3}$) used in Run6 is in the same order of magnitude as that (2.2$\times10^{-17}$\;kg m$^{-3}$) estimated for this event by employing observational techniques in \citet{2021Mannuela}. To further assess this particular run, the in-situ results obtained from Run6 are compared with the corresponding observations from the WIND spacecraft for all components of the magnetic field vector as well as the plasma density and speed (right panel of Figure~\ref{in_situ_compare1}). Apart from the magnetic field strength, the modelled speed profile (right panel, last row) also shows a remarkably good agreement with that of the observed ICME and the preceding solar wind stream. Notably, the consistency between the observed and modelled arrival time  of the high-speed stream ahead of the ICME was achieved by using the optimized rotation angle ($\psi$) for the background solar wind solution as discussed in section~\ref{bkg}.

It is important to note that the magnetic field structure of the spheromak for this particular run undergoes a rotation during its propagation such that both the northward and southward field (in HEEQ) portions (indicated by the red and blue domains) are intersected by the virtual spacecraft at Earth (see the left panel of Figure~\ref{in_situ_compare1}). However, only the magnetic axis orientation in the northward field portion of the spheromak is consistent with the axis orientation of the associated near-Sun flux rope as discussed in section~\ref{section_expansion}. The two regions enclosed by the blue and red dashed lines in the left panel of Figure~\ref{in_situ_compare1} clearly depict that the rotation of the horizontal field ($B_x \hat{x}_{HEEQ}+B_y \hat{y}_{HEEQ}$) component (indicated by the white arrows in the left panel of the figure), as well as the $B_z$ component in those two regions, are in different directions. Indeed, these two regions in the cross-sectional plane of the spheromak resemble the cross-section of two high-inclination cylindrical-like flux-ropes with oppositely directed axial magnetic fields. This indicates that in such a scenario the in-situ spacecraft encounters a double flux-rope signature when the whole spheromak passes through it. Therefore,  we only take into account the northward field portion of the spheromak that is constrained from the observations while comparing the model results with in-situ observations. In this case, the domain enclosed by the red dashed line (as depicted in the left panel of Figure~\ref{in_situ_compare1}) is part of the spheromak that bears the flux-rope signature constrained from the observations. The grey shaded region in the right panel of Figure~\ref{in_situ_compare1} indicates the temporal passage of the above-mentioned part of the spheromak where the $B_z$ component is throughout positive. Therefore, we compare the magnetic vector profiles (red solid lines) of the spheromak inside the grey shaded region with those (black solid lines) observed inside the magnetic cloud boundary as indicated by the two vertical blue dashed lines (selected based on the ICME plasma parameters as reported in \citet{Vemareddy_2015} for this event). This approach helps us to avoid any misinterpretation in the in-situ assessment of the modelled magnetic vectors due to the presence of double flux-rope signatures inside a spheromak. 

A smooth rotation from positive to negative as observed in the $B_y$ component and a predominant positive $B_z$ profile of the observed magnetic cloud (see the black solid curves in $B_y$ and $B_z$ plot within the vertical blue dashed lines in the right panel of Figure~\ref{in_situ_compare1}) are well captured by the model results within the grey shaded region. Notably, the leading front of the spheromak arrives earlier than that of the observed magnetic cloud. Therefore, for the sake of comparison, we shift the modelled magnetic vectors by 9 hours 40 minutes so that the front edge of both modelled and observed flux-rope temporally coincide. The time-shifted modelled magnetic profiles of $B_y$ and $B_z$ as indicated by the cyan solid lines depict a good agreement between the observation and model output. In particular, the modelled $B_y$ component shows excellent correspondence with the observed profile, while a decent match is achieved for the $B_z$ component. However, the modelled $B_x$ component turns out to be opposite to the observed one as the virtual spacecraft at Earth intersects through a different part of the spheromak. In absence of any significant rotation, we show in Figures~\ref{unc_1} and \ref{unc_2} that all the three components ($B_x$, $B_y$ and $B_z$) of the flux-rope magnetic field can be well captured by the model within the uncertainty limit in obtaining the CME's direction of propagation. Notably, the amount of rotation exhibited by the spheromak is the least for the high-density case (Run7) and maximum for the low-density case (Run1) (see Figure~\ref{rotation_comparison}) as compared to the other runs performed in this work. Therefore to asses how the spheromak rotation affects the uncertainty in predicting the magnetic vectors of ICMEs, we present a comparative uncertainty analysis for these two extreme cases (Run1 and Run7) in Figures~\ref{unc_1} and \ref{unc_2}.     

The in-situ assessment made for the modelled results obtained from different virtual spacecraft placed within $\pm$ 15$^{\circ}$ longitude (see Figure~\ref{unc_1}) and latitude (see Figure~\ref{unc_2}) gives insight on how the spheromak rotation affects the $B_z$ prediction at 1~au. In presence of significant rotation of the spheromak, the modelled $B_z$ component becomes highly sensitive to the spacecraft location as shown in Panel (b) of both Figures~\ref{unc_1} and \ref{unc_2}. Indeed, the figures show that completely opposite profiles of $B_z$ can be obtained at nearby spacecraft as the set of modelled $B_z$ in such a scenario can start with positive as well as negative values as depicted in Panel (b). This indicates that the spheromak rotation leads to a large uncertainty in $B_z$ prediction for this event.

On the other hand, in the absence of significant spheromak rotation, the uncertainties in $B_z$ prediction significantly reduces as shown in Panel (d) of Figures~\ref{unc_1} and \ref{unc_2}. All the possible $B_z$ profiles at different nearby virtual spacecraft shows predominant positive values of $B_z$ which is in agreement with the observed $B_z$ profile at 1~au. These results suggest that the prediction efficacy improves when a spheromak undergoes minimal rotation effect.

We also note that the simulation outputs of EUHFORIA capture well the formation of the shock as well as the sheath structure ahead of the spheromak as illustrated in Figure~\ref{sheath}. However, a detail study of the sheath region requires a higher spatial resolution in the simulation which is out of the scope of the current study. In future work, we plan to explore the formation and evolution of sheath regions in detail.     

\section{Discussion and Conclusion}\label{section5}
In this work, we have assessed the applicability of spheromaks to be used as a model for the magnetic field of CMEs in global MHD simulations for space weather forecasting. In order to test the performance of a spheromak model in real-event forecasting,  we carry out a set of data constrained MHD simulations with EUHFORIA for an Earth-impacting CME event on 2013 April 11. As the spheromaks are prone to the `tilting instability' that causes inherent rotation of its magnetic axis \citep{Asvestari2}, we perform a quantitative analysis of the resultant rotation experienced by the spheromak to understand how that affects the $B_z$ prediction at 1~au. Further, we study the role of spheromak density to mitigate the rotation of the spheromak that may reduce the uncertainty in forecasting $B_z$. The important findings of this study are discussed below in order to answer the key scientific questions as mentioned in Section~\ref{sec:intro}.
\begin{itemize} 

\item Our simulation results confirm the presence of spheromak rotation, reminiscent of the tilting instability in data-constrained MHD simulations. Indeed, we find that a significant rotation up to 90$^{\circ}$ may occur during the heliospheric propagation of a spheromak. The 3D visualisation of the spheromak rotation reveals that this rotational motion is different from the CME rotation which may occur in the lower corona due to the un-writhing motion \citep{2009ApJLynch,Zhou2022}. During the lower coronal evolution, the magnetic axis of a CME may rotate about its direction of rise or propagation \citep{Zhou2022}. In contrast, we find that the rotation of the spheromak magnetic axis as observed for this event occurs approximately about the line perpendicular to the direction of propagation. Moreover, the sense of CME rotation in the lower corona follows its chirality, i. e., CMEs with negative/positive chirality rotate anti-clockwise/clockwise \citep[][]{2009ApJLynch,Green2007,Zhou2022}. However, the sense of rotation of the magnetic axis of the spheromak is not solely dependent on its chirality but also depends on the direction of the ambient magnetic field \citep{Asvestari2}.    

\item The most significant part of the spheromak rotation is observed to take place close to the Sun below 0.3~au. The underlying reason behind this is that the ambient magnetic field is stronger close to the Sun. Therefore, the torque force exerted on the spheromak becomes less effective above 0.3~au.     

\item In the presence of significant spheromak rotation, the predicted magnetic field topology of the ICME at 1~au is expected to show poor results when the simulation output is compared with the observational signatures. Interestingly, we find that the spheromak density has a major role to mitigate its rotation effect. Running a set of simulations by using different density values within the observed range, we find that the spheromak rotates less when the density is higher. This can be explained as a consequence of the increased moment of inertia ($I$) of the more dense spheromaks. Assuming the torque ($T$) exerted by the background wind remains the same, the angular acceleration ($\alpha$) of the spheromak needs to decrease since, $T=I\times \alpha$. 

Notably, the insertion speed of the spheromak also has a role \citep{Asvestari2} on the amount of rotation experienced by the spheromak. Therefore, a similar effect could probably be achieved by not changing the density but instead the speed. However, density is the focus in the paper as in general the speed of the CME is much more well constrained from remote observations than the  mass density.

\item As the degrees of spheromak rotation significantly varies between high and low density spheromaks, different portions of the spheromak are being probed by a virtual spacecraft at Earth under varying density conditions. For the low density spheromak that exhibits the maximum rotation ($\approx 90^{\circ}$), the virtual spacecraft at Earth approximately crosses through the symmetry axis of the spheromak, thereby missing the twisted flux-rope part. On the other hand, in the case of high-density spheromaks with minimal rotation, the virtual spacecraft at Earth predominantly passes through the twisted part within the spheromak. Therefore, spheromaks with different densities lead to different in-situ magnetic profiles at Earth. In addition to this, the part of the spheromak accounted for modeling the observed flux-rope undergoes significant compression in the high-density case as compared to the low-density scenario, leading to further changes in the magnetic strength of the spheromak under varying density conditions. As a combined result of the two aforementioned phenomena, the high-density spheromaks exhibit higher field strength as probed by the virtual spacecraft at Earth.

%We also find that the high-density spheromaks undergo significant compression at the front as compared to the low-density ones. Therefore, although the rotation effect is minimized for high-density spheromaks, the associated field strength at 1~au is overestimated due to the compression effect.  

\item Our assessment of the simulation results with the in-situ observations for the event under study show that the spheromak rotation leads to large uncertainties in $B_z$ predictions, whereas the prediction efficacy significantly improves in the absence of any significant rotation of the spheromak. 
\end{itemize}

Our results imply that a different part of the spheromak than that constrained from the observations may arrive at 1~au due to its rotation in interplanetary domain. As a consequence of this rotation, the prediction of CME magnetic vectors at 1~au can be largely affected. Therefore, care must be taken when using the spheromak in global MHD models for space weather forecasting. In particular, the magnetic configuration of the simulated flux-rope should be visualised in 3D to check the effect that the changes in tilt during the propagation has on its final orientation at 1~au.

Intuitively, inserting only half of the spheromak in the heliospheric MHD domain may result in a different scenario as compared to that reported in this study. In such a scenario, when the peripheral magnetic field of the partially inserted spheromak gets peeled off due to erosion, it may mimic a flux-rope structure with two legs attached to the inner boundary of the simulation. Similar situation would be expected to arise in case of inserting a half-torus like flux-rope. Under those circumstances, the rotation of the flux-rope magnetic axis may not be as significant as that for a fully inserted spheromak or a torus. However, as claimed in \cite{Asvestari2}, a flux-rope with two legs attached to the inner boundary may still have a magnetic moment that will try to align with that of the ambient magnetic field. As the legs of the flux-rope remain fixed on the inner boundary, the manifestation of the force acting on the flux-rope under that circumstance may mostly result in a deflection in contrast to the large unrealistic rotation as expected for the cases of fully inserted spheromaks. Possibly, such deflections could be realistic and therefore would be important to incorporate in global MHD models in the perspective of space-weather forecasting. As the ambient magnetic field is largely effective below 0.3~au (see sub-section~\ref{section_rotation}), the MHD modeling approaches that superimpose a flux-rope in background wind when the leading-edge height is already at approximately 0.3~au (e.g. see \cite{Singh_2022}), are not capable of capturing such effect happening within 0.1 to 0.3~au. Therefore, we emphasize that in contrast to the method of superposing a flux-rope, the method of inserting that from the inner-boundary of the MHD model is capable of capturing its early evolution close to the inner boundary (0.1~au) of the heliospheric domain. As a future work, we plan to implement the insertion of a half-torus-like flux-rope with two legs are attached to the inner boundary of the heliospheric model in EUHFORIA. This would allow us to explore if such an implementation technique still leads to some amount of deflection and rotation of the flux-rope which can then be confirmed as the realistic scenario present in the CME evolution.

This research has received funding from the European Union’s Horizon 2020 research and innovation program under grant agreement No 870405 (EUHFORIA 2.0). RS acknowledges support from the project EFESIS (Exploring the Formation, Evolution and Space-weather Impact of Sheath-regions), under the Academy of Finland Grant 350015. JP acknowledges Academy of Finland Project 343581. EA acknowledges support from the Academy of Finland (Postdoctoral Researcher Grant 322455). EK, EA and JP acknowledge the ERC under the European Union's Horizon 2020 Research and Innovation Programme Project 724391 (SolMAG). The work at the University of Helsinki was performed under the umbrella of Finnish Centre of Excellence in Research of Sustainable Space (Academy of Finland Grant no. 312390, 336807). NW acknowledges support from NASA program NNH17ZDA001N-LWS and from the Research Foundation - Flanders (FWO-Vlaanderen, fellowship no.\ 1184319N). AM and SP acknowledge support from the projects C14/19/089 (C1 project Internal Funds KU Leuven), G.0025.23N (WEAVE FWO-Vlaanderen), SIDC Data Exploitation (ESA Prodex-12), and Belspo project B2/191/P1/SWiM. Open access is funded by Helsinki University Library.\\

\appendix
  
\begin{figure}[b!]
\begin{center}
\includegraphics[width=\textwidth,clip=]{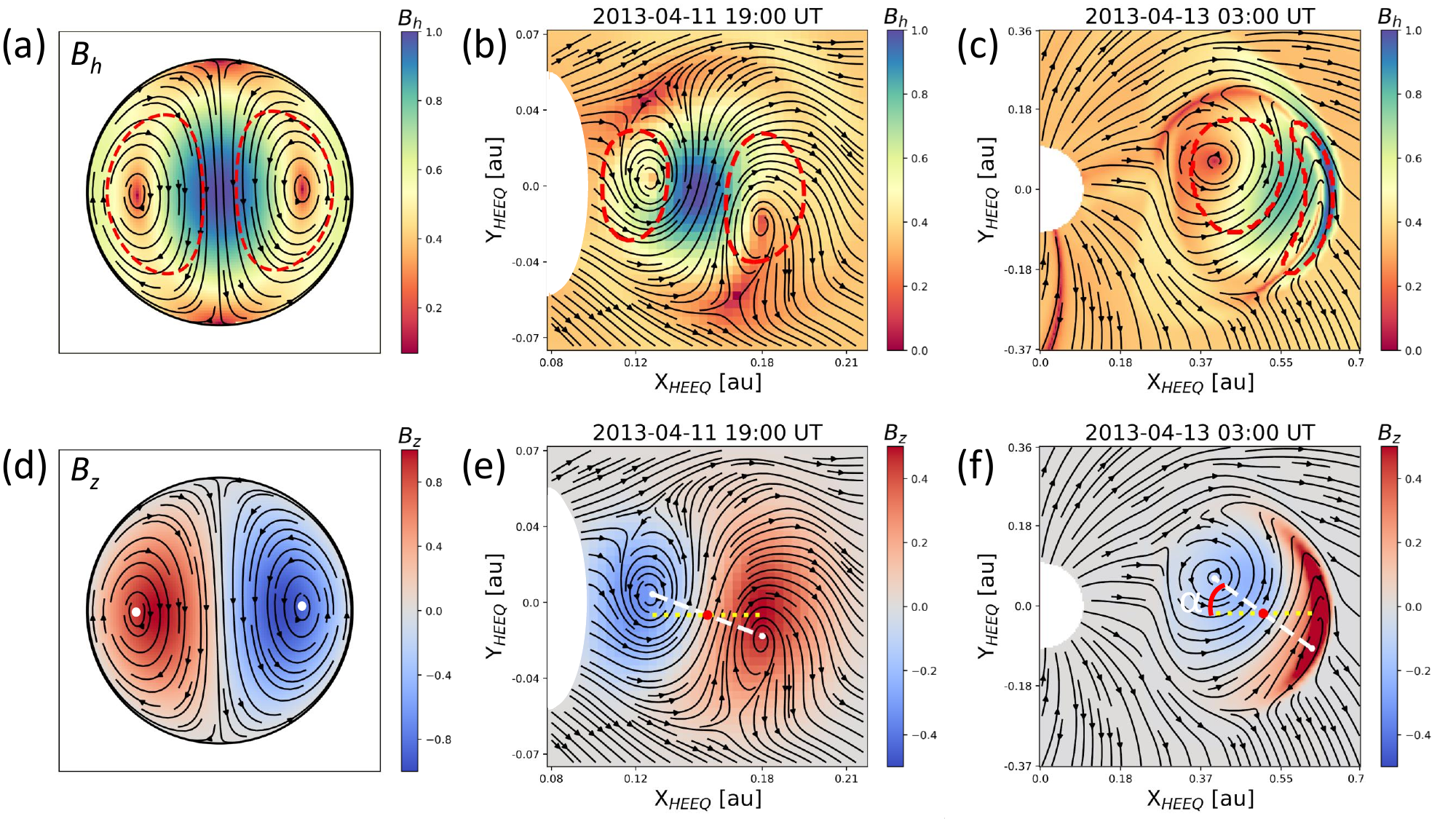}
\caption{The poloidal (panel [a]) and toroidal (panel [d]) field strength in a cross-sectional plane perpendicular to the magnetic axis of an analytical linear force free spheromak. The poloidal field is denoted by $B_h$ which lies in the plane of the image
and the toroidal field is denoted by $B_z$ which is perpendicular to the plane of the image. The black streamlines mark the poloidal field direction. The red dashed contours in panel [a] encloses the region where the strength of $B_z$ is higher than 0.6 times of the maximum $B_z$ on the cross-sectional plane shown in the figure. The two white dots in panel [d] mark the two locations where the magnetic axis passes through the plane.  The magnetic field of the spheromak on the HEEQ equatorial plane, during one of its initial (panels [b] and [e]) and later (panels [c] and [f]) phases of heliospheric propagation as obtained from Run7.
\label{rotation_appendix}}
\end{center}
\end{figure}
We estimate the trend of spheromak rotation for different density runs as shown in Figure~\ref{rotation_comparison} by identifying the orientation of its magnetic axis curve during different phases of its evolution in the heliospheric domain. For this purpose, we developed a method to estimate the projected orientation of the magnetic axis curve of a spheromak on any plane that passes close (well within the radius of the magnetic axis) to its centroid and is perpendicular to its magnetic axis. 

In order to describe the method, we present an example of a cross-sectional plane of the analytical linear force free spheromak used in EUHFORIA as shown in Figure~\ref{rotation_appendix} [a] and [d]. The spheromak solution on this cross-sectional plane is expressed in Cartesian coordinates ($B_x$, $B_y$ and $B_z$) where $B_x$ and $B_y$ lie in the plane of the figure and $B_z$ is perpendicular to the depicted plane. The strength of $B_h$ ($=\sqrt{B_x^2+B_y^2}$) as shown in Figure~\ref{rotation_appendix} [a] depicts the strength of the poloidal magnetic field, whereas the $B_z$ map as shown in Figure~\ref{rotation_appendix} [d] represents the toroidal field. The over-plotted black streamlines delineate the direction of the poloidal magnetic field which swirl around a common center (marked by the white dots in Figure~\ref{rotation_appendix} [d]) indicating the location of the magnetic axis as it crosses through the cross-sectional plane. Therefore, identifying the locations of those two swirling centers of poloidal field would allow us to estimate the projected orientation of the magnetic axis curve by connecting those two points on the aforementioned plane. The $B_h$ map as shown in Figure~\ref{rotation_appendix} [a] shows that the value of $B_h$ attains a minima at the two swirling centers of the poloidal field as indicated by the color map. Notably, it can be further observed that the $B_h$ value also decreases towards the two poles of the symmetry axis of the spheromak. Therefore, we apply a mask on the $B_h$ map as indicated by the regions enclosed by the red dashed contours (see Figure~\ref{rotation_appendix} [a]) within which the strength of $B_z$ is greater than 0.6 times the maximum strength of $B_z$ within that plane. This selection of mask helps us to discard the regions of lower $B_h$ value towards the poles of the symmetry axis. Therefore, within the masking area, identifying the two points where $B_h$ value becomes minimum, gives us the locations of the two points where the magnetic axis of the spheromak passes through the plane. Finally, connecting those two points with a straight line gives us the orientation of the projected magnetic axis curve of the spheromak. 

We apply this method to track the change in rotation angle of the spheromak magnetic axis in the simulation outputs obtained from EUHFORIA. As a first step, we identify the plane in which the magnetic axis curve of the spheromak is clearly seen to rotate during its heliospheric evolution. Based on the 3D visualisation of the spheromak magnetic axis as shown in Figure~\ref{spheromak_rotation} and\ref{rotation_3d}, we identify that the rotation of the spheromak for the runs performed in this work is well seen in the HEEQ equatorial plane. Therefore, we use the 2d map of $B_x$, $B_y$ and $B_z$ in HEEQ on the equatorial plane to track the orientation of the magnetic axis curve of the spheromak. Before applying the method, we multiply the components of the magnetic field at each grid point of the 2d map with $r^2$, where $r$ is the distance of each grid point from the Sun-center. This helps us to remove the gradient of magnetic field strength that decreases with larger helio distances ($r$) following the $\frac{1}{r^2}$ relation. 

We show the results of our tracking method during two different time-frames for Run7 as shown in Figure~\ref{rotation_appendix} [e] and [f]. Figure~\ref{rotation_appendix} [b] and [c] show the HEEQ $B_h$ ($=\sqrt{B_x^2+B_y^2}$) component of the spheromak magnetic field at the equatorial plane. The over-plotted red contours on the aforementioned $B_h$ maps enclose the masking area within which we identify the local minima of $B_h$. We further connect the identified conjugate locations of the $B_h$ minima by a white dashed line as plotted on top of the $B_z$ maps shown in Figure~\ref{rotation_appendix} [e] and [f]. This white dashed line represents the projected magnetic axis curve as identified from the tracking method. Comparing this with the background streamline plot of the equatorial component of the magnetic field, shows that the locations of two swirling points of poloidal magnetic field inside the spheromak cross-section are correctly identified by the two endpoints of the white-dashed line. We consider the centroid (as indicated by the red dot) of the two end points of the projected magnetic axis curve as the reference point to measure the distance of the spheromak from the Sun center. The rotation angle ($\alpha$) of the projected magnetic axis curve as plotted in Figure~\ref{rotation_comparison} is measured with respect to the reference line (parallel to the $X_{HEEQ}$ or Sun-Earth line) as indicated by the yellow dashed line in Figure~\ref{rotation_appendix} [f]. 

Notably, the tracking method applied in this work is different from the one employed in \citet{Asvestari2} which tracks the symmetry axis instead of the magnetic axis of the spheromak to estimate its orientation angle. However, tracking the magnetic axis as demonstrated in this work turns out to be a useful alternative tool incorporating the cases of high density runs (e.g. Run7) where it becomes difficult to define the symmetry axis due to the high compression at the frontal part of the spheromak (see Figure~\ref{density_compression} [f] and \ref{rotation_appendix} [f]).  
\end{document}